\documentclass[12pt]{article}
\pdfoutput=1

\usepackage{putex}
\usepackage{graphicx}
\usepackage{caption}
\usepackage{amsmath}
\usepackage{amssymb}
\usepackage{array}
\usepackage{multirow}
\usepackage{mathtools}
\usepackage{comment}
\usepackage{subcaption}
\usepackage{epstopdf}
\usepackage{enumerate}
\usepackage{cite}
\usepackage{youngtab}
\usepackage{tensor}
\usepackage{slashed}
\usepackage[aligntableaux=center]{ytableau}
\usepackage[utf8]{inputenc}
\usepackage{rotating}
\usepackage{bigfoot}
\usepackage[
      colorlinks=true,
      linkcolor=blue,
      urlcolor=blue,
      filecolor=black,
      citecolor=red,
      linktocpage=true
      ]{hyperref}

\newcommand{\ec}{\,,}
\newcommand{\ecq}{\ec\quad}

\newcommand{\abs}[1]{\left\lvert #1 \right\rvert}
\def\Tr{\mop{Tr}}
\newcommand {\be} {\begin {equation}}
\newcommand {\ee} {\end {equation}}

\newcommand {\bes} {\begin {equation*}}
\newcommand {\ees} {\end {equation*}}

\newcommand{\es}[2] {\begin{equation} \label{#1} \begin{split} #2 \end{split} \end{equation}}

\newcommand{\cF}{{\mathcal F}}
\newcommand{\cG}{{\mathcal G}}

\newcommand{\cS}{{\mathcal S}}
\newcommand{\cT}{{\mathcal T}}

\newcommand{\bea}{\begin{equation}\begin{aligned}}
\newcommand{\eea}[1]{\label{#1}\end{aligned}\end{equation}}

\newcommand{\beq}{\begin{equation}}
\newcommand{\eeq}{\end{equation}}

\def\ie{\begin{equation}\begin{aligned}}
\def\fe{\end{aligned}\end{equation}}

\numberwithin{equation}{section}

\def\<{\langle}
\def\>{\rangle}

\begin{document}

\preprint{}

\institution{Exile}{Department of Particle Physics and Astrophysics, Weizmann Institute of Science, Rehovot, Israel}

\title{Bootstrapping 4d $\mathcal{N}=2$ gauge theories: the case of SQCD}

\authors{
Shai M. Chester\worksat{\Exile}\footnote{e-mail: {\tt iahs81@gmail.com}} 
}

\abstract{
We derive exact relations between certain integrals of the conserved flavor current four point function in 4d $\mathcal{N}=2$ conformal field theories (CFTs) and derivatives of the mass deformed sphere free energy, which can be computed exactly for gauge theories using supersymmetric localization. For conformal gauge theories with flavor groups of rank greater than one, there are at least two such integrated constraints, which can then be combined with the numerical conformal bootstrap to bound CFT data as a function of the complexified gauge coupling $\tau$. We apply this strategy to the case of $SU(2)$ conformal SQCD with flavor group $SO(8)$, where we compute bounds on unprotected scaling dimensions as a function of $\tau$ that match the free theory limit, and exhibit the expected mixing between the action of the $SL(2,\mathbb{Z})$ duality group and $SO(8)$ triality.
}
\date{}

\maketitle

\tableofcontents

\section{Introduction}
\label{intro}

Gauge theories in four spacetime dimensions are the most physical quantum field theories, as they describe the world we live in. They are also the easiest to study when the gauge coupling is small, in which case the theory is weakly coupled and physical observables can be computed using Feynman diagrams. At strong coupling, however, these theories are very difficult to study, as they require non-perturbative methods. For the most symmetric 4d gauge theory, $\mathcal{N}=4$ Super-Yang-Mills (SYM) with $N$ colors, the theory becomes integrable in the leading large $N$ limit \cite{Beisert:2010jr}, so observables can be computed at any coupling. For gauge theories with less supersymmetry in 4d, however, there is no evidence yet for integrability,\footnote{See \cite{Gadde:2010zi,Pomoni:2019oib,Pomoni:2021pbj} for discussion of this question for $SU(N)$ SQCD in the Veneziano limit.}\footnote{Integrability can also be applied to the planar limit of a certain non-unitary non-gauge theory called the fishnet theory, obtained from taking a certain double scaled limit of $\mathcal{N}=4$ SYM and throwing out the gauge fields \cite{Gurdogan:2015csr}.} so other non-perturbative methods are required. 

The conformal bootstrap and supersymmetric localization are two other non-perturbative methods that can be used to study strongly coupled gauge theories. Localization can be applied to 4d gauge theories with at least $\mathcal{N}=2$ supersymmetry to compute protected non-local observables like the mass deformed sphere free energy and Wilson loops  \cite{Pestun:2007rz,Pestun:2016zxk}, as well as the protected OPE coefficients of certain chiral operators \cite{Baggio:2014ioa,Baggio:2014sna,Baggio:2015vxa}. However, localization cannot compute unprotected observables. The conformal bootstrap can be used to place numerical bounds on any (even unprotected) local observable in a 4d gauge theory with conformal symmetry \cite{Beem:2013qxa,Beem:2016wfs,Alday:2013opa,Bissi:2020jve,Alday:2021vfb,Beem:2014zpa,Lemos:2015awa,Gimenez-Grau:2020jrx},\footnote{See also \cite{Bissi:2021rei,Cornagliotto:2017snu} for bootstrap bounds on 4d $\mathcal{N}=2$ non-Lagrangian theories, \cite{Poland:2018epd,Simmons-Duffin:2016gjk,Chester:2019wfx,Poland:2022qrs} for general reviews on the numerical bootstrap, and \cite{Rattazzi:2008pe} for the original numerical bootstrap paper.} but these bounds are a priori only sensitive to the global symmetries of the theory, and so cannot be computed as a function of the coupling. All 4d $\mathcal{N}=2$ conformal gauge theories are conformal manifolds with at least one complex parameter, due to the complexified gauge coupling(s) $\tau$, so one needs to input at least two exactly known quantities that are functions of $\tau$ to pin down the theory.\footnote{While in principle infinite protected OPE coefficients can be computed using the localization methods of \cite{Baggio:2014ioa,Baggio:2014sna,Baggio:2015vxa}, in practice the instanton contributions to the localization formula are only known in general for the single OPE coefficient appearing in the 4-point of the lowest dimension chiral operator \cite{Gerchkovitz:2016gxx,Fucito:2015ofa}, which is insufficient to fix a point on conformal manifold with at least one complex dimension.}

In this work we show how localization can be combined with the conformal bootstrap to compute unprotected observables in conformal 4d $\mathcal{N}=2$ gauge theories for any value of $\tau$. Following previous work on 4d $\mathcal{N}=4$ conformal field theories (CFTs) \cite{Binder:2019jwn,Chester:2020dja,Chester:2021aun}\footnote{These 4d integrated constraints were inspired by the original work on 3d integrated constraints in \cite{Binder:2018yvd,Binder:2019mpb}.}, we derive at least two independent relations\footnote{We will show that the number of relations corresponds to the number of quartic casimirs for the flavor symmetry group of the theory, which is two or more for all groups of rank greater than one. This includes all known 4d $\mathcal{N}=2$ conformal gauge theories except $\mathcal{N}=4$ SYM, when considered as an $\mathcal{N}=2$ theory, for which additional constraints are known due to the increased supersymmetry \cite{Binder:2019jwn}.} between derivatives of the mass deformed sphere free energy $F(m)$ in 4d $\mathcal{N}=2$ CFTs and certain integrals of the four point function of the conserved flavor current multiplet (often called the moment map 4-point due to its superprimary). These relations are particularly useful for gauge theories, where $F(m)$ can be computed non-perturbatively as a function of $\tau$ using localization. Since we have at least two such localization inputs, we can thus fix a point on the 2-dimensional real conformal manifold of a CFT with a simple gauge group.\footnote{For theories with multiple gauge groups, there will be multiple complex parameters of the conformal manifold, so one requires additional inputs to specify the theory.} The conformal bootstrap can then be applied to bound CFT data in the four point function non-perturbatively as a function of $\tau$.\footnote{See \cite{Lin:2015wcg,Baggio:2017mas} for previous work on bootstrapping conformal manifolds in 2d and 3d, respectively.}

As a case study, we consider the simplest conformal 4d $\mathcal{N}=2$ gauge theory: $SU(2)$ SQCD, which has four hypermultiplets that transform in the fundamental irrep of an enhanced $SO(8)$ flavor symmetry.\footnote{For $SU(N)$ conformal SQCD with $N>2$, there are $2N$ hypermultiplets transforming in a $U(2N)$ flavor group. The symmetry enhances only for $N=2$ because the fundamental of $SU(2)$ is pseudo-real, so we can think of the 4 hypermultiplets in the fundamental of $U(4)$ as 8 half-hypermultiplets in the fundamental of $SO(8)$.} This theory is a conformal manifold parameterized by $\tau$ with duality group $SL(2,\mathbb{Z})$, where the $SO(8)$ triality frame transforms under the action of the duality group \cite{Seiberg:1994aj}. We will derive three relations between integrals of the moment map 4-point and quartic mass derivatives of $F(m)$, whose localization expression is given by a certain 2d Liouville correlator according to the AGT correspondence \cite{Alday:2009aq}. These three mass derivatives can be associated to the ${\bf 35_c}$, ${\bf 35_s}$, and ${\bf 35_v}$ irreps of $SO(8)$ that appear in the 4-point, which are permuted by triality and thus also by $SL(2,\mathbb{Z})$ duality. We then combine these integrated constraints with the numerical conformal bootstrap as in \cite{Chester:2021aun} to compute non-perturbative bounds as a function of $\tau$ on the unprotected scaling dimensions $\Delta_{\bf 35_c}$, $\Delta_{\bf 35_s}$, and $\Delta_{\bf 35_v}$ of the lowest dimension scalar operators in these irreps. We find that our bounds approximately match the free theory expressions at zero coupling, and at strong coupling the bounds of the different triality related irreps interchange as expected from $SL(2,\mathbb{Z})$ duality. This motivates the conjecture that in the limit of infinite numerical precision, the physical theory is close to saturating these bounds, which therefore provides the first calculation of unprotected strongly coupled observables in an $\mathcal{N}=2$ gauge theory.

 The rest of this paper is organized as follows. In Section~\ref{4point} we review the constraints of superconformal symmetry on the moment map correlator for general 4d $\mathcal{N}=2$ CFTs, as well as new results for the case of $SU(2)$ SQCD with flavor group $SO(8)$.  In Section \ref{intCon}, we derive non-perturbative relations between mass derivatives of $F(m)$ and integrals of the moment map 4-point, which we then apply to $SU(2)$ SQCD, where we compute the mass derivatives explicitly for any $\tau$ using localization and the AGT correspondence. In Section \ref{intBoot}, we combine these integrated constraints with the numerical conformal bootstrap to compute non-perturbative bounds on $\Delta_{\bf 35_c}$, $\Delta_{\bf 35_s}$, and $\Delta_{\bf 35_v}$ as a function of $\tau$. We conclude in Section \ref{conc} with a review of our results and a discussion of future directions. Technical details of the localization calculation are given in the Appendix \ref{locApp}, and we include an attached \texttt{Mathematica} notebook with some explicit results.

\section{$\mathcal{N}=2$ moment map four-point function}
\label{4point}

The main object of study in this work is the four-point function of the flavor conserved current multiplet, whose lowest component is called the moment map operator. We begin by reviewing kinematic constraints on this correlator coming from invariance under the superconformal algebra for general $\mathcal{N}=2$ CFTs with flavor group $G$, following \cite{Dolan:2001tt,Beem:2014zpa}. We will then present new results for the case $G=SO(8)$, which applies to SQCD with gauge group $SU(2)$.

\subsection{General case}
\label{gen4point}

The moment map operator is a Lorentz scalar of dimension $\Delta=2$ that transforms in the adjoint of the flavor group $G$ and the $SU(2)_R$ subgroup of the R-symmetry, and is invariant under the $U(1)_R$ R-symmetry subgroup. We denote this operator as $\phi^A(y,x)$, where $y$ is an $SU(2)_R$ polarization, $A=1,\dots,\dim(G)$ is an adjoint index of $G$, and $x$ is the position. Conformal and $SU(2)_R$ symmetry restricts the 4-point to be
\es{phiExp1}{
\langle \phi^A(y_1,x_1) \phi^B(y_2,x_2) \phi^C(y_3,x_3) \phi^D(y_4,x_4) \rangle = \frac{\langle y_1,y_2\rangle^2 \langle y_3,y_4\rangle^2}{x_{12}^4x_{34}^4}G^{ABCD}(U,V;w)\,,
}
where we define the cross ratios
\es{w}{
  U \equiv \frac{{x}_{12}^2 {x}_{34}^2}{{x}_{13}^2 {x}_{24}^2} \,, \qquad
   V \equiv \frac{{x}_{14}^2 {x}_{23}^2}{{x}_{13}^2 {x}_{24}^2}  \,, \qquad w=\frac{\langle y_1,y_2\rangle \langle y_3,y_4\rangle}{\langle y_1,y_3\rangle \langle y_2,y_4\rangle}\,, 
}
with $x_{12}\equiv x_1-x_2$ and $\langle y_1,y_2\rangle=y^\alpha_1y^\beta_2\varepsilon_{\alpha\beta}$ for $\alpha,\beta=1,2$. We can furthermore impose the flavor symmetry by expanding in projectors $P_r^{ABCD}$ for each flavor irrep $r$ as
\es{flavorProj}{
G^{ABCD}(U,V;w)=\sum_{r\in\text{Adj}\otimes\text{Adj}}G_r(U,V;w)P_r^{ABCD}\,,
}
where the projectors are normalized so that
\es{projNorm}{
P_r^{ABBA}=\text{dim}(R_r)\,.
}
The last kinematic constraint comes from the superconformal Ward identity, which we can formally solve by writing $G^{ABCD}(U,V;w)$ as \cite{Dolan:2001tt}
\es{ward4dN2}{
G_{r}(U,V;w)=\frac{z(w-\bar z)f_{r}(\bar z)-\bar z(w-z)f_{r}(z)}{w(z-\bar z)}+(1-\frac zw)(1-\frac{\bar z}{w})\mathcal{G}_{r}(U,V)\,,
}
where $U=z\bar z\,, V=(1-z)(1-\bar z)$ and the reduced correlator $\mathcal{G}$ as well as the holomorphic correlator $f(z)$ are now R-symmetry singlets. Since the holomorphic correlator $f(z)$ is protected, we can compute it from the free theory. The full free theory correlator can be computed using Wick contractions to get
\es{freeSO8}{
G_\text{free}^{ABCD}(U,V;w)&=\delta^{AB}\delta^{CD}+ \frac{U^2}{w^2}\delta^{AC}\delta^{BD}+(w^{-1}-1)^2\frac{U^2}{V^2} \delta^{AD}\delta^{BC}+\frac{2U}{kW}\Tr(T^AT^BT^DT^C)\\
 &\quad+(1-w^{-1})\frac{2U}{kV}\Tr(T^AT^BT^CT^D)+\frac{2U^2(w^{-1}-1)}{kVw}\Tr(T^AT^DT^BT^C)\,,\\
}
where the structure constants are defined in terms of the generators as
\es{fGen}{
[T^A,T^B]=iT^Cf^{ABC}\,,
}
the flavor central charge $k$ is defined in terms of the canonically normalized current 2-point of the flavor current $J_\mu^A$ as
\es{kDef}{
J_\mu^A(x) J_\nu^B(0)=\frac{3k\delta^{AB}}{4\pi^4}\frac{x^2\delta_{\mu\nu}-2x_\mu x_\nu}{x^8}\,,
}
and we normalized the generators $T$ such that the length of the largest root squared is $2$. We can then expand the free theory correlator as \eqref{ward4dN2} to get \cite{Beem:2013sza,Beem:2014zpa}
\es{freeGen}{
f^{ABCD}(z)=\delta^{AB}\delta^{CD}+z^2\delta^{AC}\delta^{BD}+\frac{z^2}{(1-z)^2}\delta^{AD}\delta^{CB}+\frac{2z}{k}f^{ACE}f^{BDE}+\frac{2z}{k(z-1)}f^{ADE}f^{BCE}\,,
}
which holds even for general interacting CFTs.

By taking the OPE twice, we can expand $f$ and $\mathcal{G}$ in terms of conformal blocks as \cite{Beem:2014zpa}
\es{blockExpSQCD}{
\mathcal{G}_r(U,V)&=U^{-1}\sum_{\ell}\sum_{\Delta\geq\ell+2}\lambda^2_{\Delta,\ell,r}G_{\Delta+2,\ell}(U,V)+U^{-1}\sum_{\ell}\lambda_{\ell+5,\ell+1,r}^2G_{\ell+5,\ell+1}(U,V)+\lambda_{4,0,r}^2G_{4,0}(U,V)\,,\\
f_r(z)&=\sum_{\ell}\lambda_{\ell+5,\ell+1,r}^2k_{2\ell+6}(z)+\lambda_{4,0,r}^2k_{4}(z)+\delta_{r,{\bf 1}}\dim(G)-\delta_{r,\text{Adj}}\frac{4h^{\vee}}{k}k_2(z)-\delta_{r,{\bf1}}\frac{\text{dim}(G)}{6c}k_4(z)\,,\\
}
where we have even/odd spin $\ell$ blocks (keeping in mind possible shifts in spin for the protected multiplets) for the irreps $r$ in the symmetric/antisymmetric product of the adjoint, and the 4d blocks and lightcone blocks are defined as
\es{4dblock}{
 G_{\Delta,\ell}(U,V) &=\frac{z\bar z}{z-\bar z}(k_{\Delta+\ell}(z)k_{\Delta-\ell-2}(\bar z)-k_{\Delta+\ell}(\bar z)k_{\Delta-\ell-2}( z))\,,\\
k_h(z)&\equiv z^{\frac h2}{}_2F_1(h/2,h/2,h,z)\,.
}
In the holomorphic block expansion we have the identity, flavor conserved current, and stress tensor multiplets, as well as the twist 4 short multiplets that also appear in the long multiplet. We normalize the coefficient of the identity block to be $\dim(G)$, while the coefficients of the conserved currents are fixed by conformal Ward identities \cite{Osborn:1993cr} in terms of the flavor central charge $k$, the conformal anomaly $c$, and the dual Coxeter number $h^{\vee}$. We can use the known holomorphic correlator in \eqref{freeGen} to also fix the protected OPE coefficients $\lambda_{\ell+5,\ell+1,r}$ and $\lambda_{4,0,r}$, so that the only undetermined data will be the long multiplet CFT data. To do this, we must now specify to a given $G$.

\subsection{SQCD }
\label{SQCD4point}

We will now illustrate the general formulae of the previous section for SQCD with gauge group $SU(2)$, which has flavor symmetry $G=SO(8)$ and central charges \cite{Beem:2013sza} 
\es{kcSO8}{
k=4\,,\qquad c=\frac76\,.
}
The adjoint transforms in the ${\bf 28}$, so we consider the tensor product:
\es{tensso8}{
{\bf28}\otimes{\bf28}={\bf1}\oplus {\bf28}\oplus {\bf35_v}\oplus {\bf35_c}\oplus {\bf35_s}\oplus {\bf300}\oplus {\bf350}\,,
}
where $SO(8)$ triality permutes the three 35-dimensional irreps. To compute the 7 tensor structures, we use the basis\footnote{Note that the last structure is unique to $SO(8)$. It would not exist for $O(8)$, in which case there would be just 6 projectors, as is the case for the decomposition of adjoints of $SO(N)$ for $N>8$.} 
\es{SO8basis}{
&\Big(\delta^{AB}\delta^{CD}\,,\quad \delta^{AC}\delta^{BD}\,,\quad \delta^{AD}\delta^{BC}\,,\quad f^{ACE}f^{BDE}\,,\quad f^{ADE}f^{BCE}\,,\\
&\qquad \Tr(T^AT^BT^CT^D)\,,\quad\frac{\varepsilon_{a_1a_2b_bj_2c_1c_2d_1d_2}}{4\cdot4!}T_{a_1a_2}^AT_{b_1b_2}^BT_{c_1c_2}^CT_{d_1d_2}^D\Big)\,,
}
where $a,b,c,d$ are fundamental indices and the generators are normalized as
\es{T}{
\Tr(T_A,T_B)=2\delta_{AB}\,.
}
The projectors \eqref{flavorProj} were computed in \cite{Isaev:2020kwc}, and in our basis take the form
\es{SO8Proj}{
P_{\bf1}&=\begin{pmatrix} \frac{1}{28}&0&0&0&0&0&0\end{pmatrix}\,,\\
P_{\bf28}&=\begin{pmatrix} 0&0&0&-\frac{1}{12}&\frac{1}{12}&0&0\end{pmatrix}\,,\\
P_{\bf35_c}&=\begin{pmatrix} 0&\frac{1}{12}&\frac{1}{12}&0&\frac{1}{12}&-\frac{1}{12}&\frac12\end{pmatrix}\,,\\
P_{\bf35_s}&=\begin{pmatrix} 0&\frac{1}{12}&\frac{1}{12}&0&\frac{1}{12}&-\frac{1}{12}&-\frac12\end{pmatrix}\,,\\
P_{\bf35_v}&=\begin{pmatrix} -\frac{1}{12}&0&0&\frac{1}{12}&-\frac{1}{12}&\frac16&0\end{pmatrix}\,,\\
P_{\bf350}&=\begin{pmatrix} 0&-\frac12&\frac12&\frac{1}{12}&-\frac{1}{12}&0&0\end{pmatrix}\,,\\
P_{\bf300}&=\begin{pmatrix} \frac{1}{21}&\frac13&\frac13&-\frac{1}{12}&-\frac{1}{12}&0&0\end{pmatrix}\,.\\
}
We can use these projectors to compute the protected OPE coefficients squared appearing in $f_r(z)$ using \eqref{freeGen} to get
\es{SO8OPE}{
\lambda^2_{\ell+5,\ell+1,{\bf1}}&=
\frac{\sqrt{\pi } 2^{-2 \ell-3} ((\ell+2) (\ell+3) k-24) \Gamma (\ell+3)}{k \Gamma
   \left(\ell+\frac{5}{2}\right)}
\,,\\
   \lambda^2_{\ell+5,\ell+1,{\bf28}}&=\frac{\sqrt{\pi } 2^{-2 \ell-3} ((\ell+2) (\ell+3) k-12) \Gamma (\ell+3)}{k \Gamma
   \left(\ell+\frac{5}{2}\right)}\,,\\
   \lambda^2_{\ell+5,\ell+1,{\bf35_s}}&=\lambda^2_{\ell+5,\ell+1,{\bf35_c}}=   \lambda^2_{\ell+5,\ell+1,{\bf35_v}}=\frac{\sqrt{\pi } 2^{-2 \ell-3} ((\ell+2) (\ell+3) k-8) \Gamma (\ell+3)}{k \Gamma
   \left(\ell+\frac{5}{2}\right)}\,,\\
   \lambda^2_{\ell+5,\ell+1,{\bf350}}&=\frac{\sqrt{\pi } 2^{-2 \ell-3} (\ell+2) \Gamma (\ell+4)}{\Gamma
   \left(\ell+\frac{5}{2}\right)}\,,\\
   \lambda^2_{\ell+5,\ell+1,{\bf300}}&=\frac{\sqrt{\pi } 2^{-2 \ell-3} ((\ell+2) (\ell+3) k+4) \Gamma (\ell+3)}{k \Gamma
   \left(\ell+\frac{5}{2}\right)}\,,\\
   \lambda^2_{4,0,{\bf1}}&=   \lambda^2_{4,0,{\bf35_v}}=   \lambda^2_{4,0,{\bf35_c}} =    \lambda^2_{4,0,{\bf35_s}} =0\,,\qquad    \lambda^2_{4,0,{\bf300}}=3\,,
}
where we see that several OPE coefficients vanish for the values of $k,c$ in \eqref{kcSO8}. As a consistency check, we also recover the known flavor multiplet coefficient in \eqref{blockExpSQCD}, where $h^{\vee}=6$ for $SO(8)$. Note that all the 35-dimensional irreps have the same coefficients. As discussed in \cite{Beem:2013sza}, this is expected because the $SO(8)$ triality frame changes under the action of $S$-duality, but short multiplets are $S$-duality invariants (since they do not depend on $\tau$). For the long multiplets, the different $35$ structures will get different values, which violates triality. For instance, the free theory correlator \eqref{freeSO8} can be written in terms of the reduced and holomorphic correlators in \eqref{ward4dN2} as
\es{freeGSO8}{
&\vec{\mathcal{G}}_\text{free}(U,V)=\Big(\frac{U\left(V^2+V+1\right)}{V^2}\quad\quad \frac{U}{V^2}-U \quad\quad \frac{U
   (V-1)^2}{V^2}\quad\quad\frac{U(V-1)^2}{V^2}\\
   &\qquad\qquad\qquad\frac{U\left(V^2+V+1\right)}{V^2}\quad\quad
   \frac{U}{V^2}-U\quad\quad\frac{U\left(V^2+V+1\right)}{V^2}\Big)\,,\\
&\vec f(z) = \Big( \frac{z^2}{(z-1)^2}+z^2+12z +28  \quad\quad \frac{2z^3-z^4}{(z-1)^2} \quad\quad 
   \frac{z^2}{(z-1)^2}+z^2+4z \\
   &\frac{z^2}{(z-1)^2}+z^2+4z  \quad\quad 
   \frac{z^2}{(z-1)^2}+z^2+4z\quad\quad \frac{2z^3-z^4}{(z-1)^2} \quad\quad
   \frac{z^2}{(z-1)^2}+z^2-2z\Big)\,,\\
}
where we write this in the irrep basis \eqref{SO8Proj}. We can see that $\mathcal{G}_\text{free}$ has a different coefficient for the ${\bf35_v}$, and so violates triality, while $f(z)$ has the same coefficient for all the triality related 35-dimensional irreps.

We can then plug the short OPE coefficients \eqref{SO8OPE} into the block expansion \eqref{blockExpSQCD} and resum so that the full block expansion can be written as
\es{blockExpSQCD2}{
\mathcal{G}_r(U,V)&=U^{-1}\sum_{\ell}\sum_{\Delta\geq\ell+2}\lambda^2_{\Delta,\ell,r}G_{\Delta+2,\ell}(U,V)+\mathcal{F}_\text{short}^{r}(U,V)\,,
}
where the short pieces are
\es{shortsSO8}{
\mathcal{F}_\text{short}^{{\bf1}}&=\frac{4 \log (1-z)}{4(z-{\bar z})} \left(48 \left(\frac{1}{z}-\frac{1}{{\bar z}}\right) \log
   (1-{\bar z})+{\bar z}
   ({\bar z}+6)+\frac{8}{{\bar z}-1}+\frac{1}{({\bar z}-1)^2}-41\right)\\
   &-\frac{4 (z (z (z (z+4)-52)+96)-48) \log (1-{\bar z})}{  4(z-{\bar z})(z-1)^2}
\,,\\
\mathcal{F}_\text{short}^{{\bf28}}&=\frac{\frac{4 z \left(z \left(z^2+z-9\right)+6\right) \log
   (1-{\bar z})}{(z-1)^2}-\frac{4 {\bar z} \left({\bar z}
   \left({\bar z}^2+{\bar z}-9\right)+6\right) \log
   (1-z)}{({\bar z}-1)^2}}{4 (z-{\bar z})}\,,\\
\mathcal{F}_\text{short}^{{\bf35_s}}&=\mathcal{F}_\text{short}^{{\bf35_c}}=\mathcal{F}_\text{short}^{{\bf35_v}}=\frac{\frac{{\bar z}^4 \log (1-z)}{({\bar z}-1)^2}-\frac{z^4 \log
   (1-{\bar z})}{(z-1)^2}}{z-{\bar z}}\,,\\
\mathcal{F}_\text{short}^{{\bf350}}&=\frac{\frac{(z-2) z^3 \log (1-{\bar z})}{(z-1)^2}-\frac{({\bar z}-2)
   {\bar z}^3 \log (1-z)}{({\bar z}-1)^2}}{z-{\bar z}}\,,\\
\mathcal{F}_\text{short}^{{\bf300}}&=\frac{\frac{{\bar z}^2 (({\bar z}-3) {\bar z}+3) \log
   (1-z)}{({\bar z}-1)^2}-\frac{z^2 ((z-3) z+3) \log
   (1-{\bar z})}{(z-1)^2}}{z-{\bar z}}\,.\\
}
The nontrivial content of the moment map 4-point function is thus given entirely by the scaling dimensions $\Delta$ and OPE coefficients $\lambda_{\Delta,\ell,r}$ of the long multiplets, which we will seek to constrain in the following sections.

\section{Integrated constraints from sphere partition function}
\label{intCon}

We will now derive exact relations between certain integrals of the moment map 4-point function and derivatives of the mass deformed sphere free energy $F(m_A)$. While these relations hold for any $\mathcal{N}=2$ CFT, they are most useful for gauge theories, in which case $F(m_A)$ can be computed non-perturbatively using supersymmetric localization. As in the previous section, we will begin by describing the general case, and then restrict to $SU(2)$ SQCD.

\subsection{General case}
\label{intConGen}

Consider the flavor multiplet $(J_{\mu},\Sigma,\phi_{(\alpha\beta)},\dots)_A$, where the bosonic operators consist of the current $J_{\mu}$, complex $\Delta=3$ scalars $\Sigma$, and the real moment map operator $\phi_{(\alpha\beta)}$, which we write here with explicit $SU(2)_R$ indices. All of these transform in the adjoint of the flavor group $G$ with index $A$. The mass deformation of the action is
\es{Sm}{
S_m=\int d^4x\sqrt{g}\Big(m_A\Big[\frac irN_J(\phi^A_{11}+\phi^A_{22})+N_K(\Sigma^A+\bar\Sigma^A)\Big]\Big)+m^2L\,,
}
where $r$ is the radius of the sphere, the quadratic term $L$ will not matter in what follows, and the operators are normalized with 2-points 
\es{normBSig}{
\langle \phi(x_1,y_1) \phi(x_2,y_2)\rangle=\frac{\langle y_1,y_2\rangle^2}{x_{12}^4}\,,\qquad \langle \Sigma(x_1)\bar\Sigma(x_2)\rangle=\frac{1}{x_{12}^6}\,.
}
The conformal Ward identity requires that the normalization factors $N_J$ and $N_K$ be proportional to the square root of the flavor central charge $k$, as we showed explicitly for the current 2-point in \eqref{kDef}. We can take four derivatives of various masses to derive the integrated constraints
 \es{FourthMass}{
    -\partial_{m_{A}} \partial_{m_{B}} \partial_{m_{C}} \partial_{m_{D}} F\big|_{m=0} &= \Bigg \langle  \left( \int d^4 \vec{x}\, \sqrt{g}\Big[\frac irN_J(\phi^A_{11}+\phi^A_{22})+N_K(\Sigma^A+\bar\Sigma^A)\Big]\right) \\
    &\qquad \cdots   \left( \int d^4 \vec{x}\, \sqrt{g}\Big[\frac irN_J(\phi^D_{11}+\phi^D_{22})+N_K(\Sigma^D+\bar\Sigma^D)\Big]\right)\Bigg\rangle\\
 & \qquad + \text{(2- and 3-pt function contributions)}\,,
 }
 where the last line involves the quadratic operator $L$, while the other terms are integrals over $S^4$ of 4-points of $\phi$ and $\Sigma$. Since the flavor current multiplet is half-BPS, all these 4-points can be uniquely written in terms of the moment map 4-point function.\footnote{We also checked this statement explicitly by deriving the Ward identities using explicit components, similar to the $\mathcal{N}=4$ case in \cite{Chester:2020dja}.} Also, the 4-point function only depends on the conformal cross ratios $U,V$, so we expect this integrated constraint can be simplified to take the form
 \es{simple}{
  -\partial_{m_{A}} \partial_{m_{B}} \partial_{m_{C}} \partial_{m_{D}} F\big|_{m=0} &= -\partial_{m_{A}} \partial_{m_{B}} \partial_{m_{C}} \partial_{m_{D}} F\big|^\text{free}_{m=0} \\
  &\qquad+k^2\int dUdV {\bf f}(U,V)\sum_{r}P_r^{(ABCD)} \mathcal{G}^\text{int}_r(U,V)\,,
  }
  where we symmetrized over flavor indices so that the RHS should be crossing invariant, we defined $\mathcal{G}^\text{int}_r(U,V)\equiv \mathcal{G}_r(U,V)-\mathcal{G}^\text{free}_r(U,V)$ by subtracting the free theory correlator \eqref{freeSO8}, and we wrote the free theory contribution (which includes the contributions from the last line of \eqref{FourthMass}) separately. Note that the dependence on the specific theory and flavor group $G$ is entirely captured by the projectors $P_r$ and the flavor central charge $k$, while $ {\bf f}(U,V)$ is a theory-independent measure fixed by superconformal symmetry. 
  
  To fix $ {\bf f}(U,V)$, we will consider the specific example of $\mathcal{N}=4$ Super-Yang-Mills, which can be thought of as an $\mathcal{N}=2$ CFT with $G=SU(2)$, where the $\mathcal{N}=4$ stress tensor multiplet decomposes to the $\mathcal{N}=2$ flavor current multiplet (as well as other multiplets). The $\mathcal{N}=4$ integrated constraint acts on the stress tensor superprimary $S$, which is a $\Delta=2$ scalar transforming in the $\bf20'$ of $SU(4)_R$, and was shown in \cite{Chester:2020dja} to take the form
  \es{constraint2}{
  -c^{-2} \partial^4_mF\big|_{m=0}   &=   -c^{-2} \partial^4_mF\big|^\text{free}_{m=0} 
   +I[\cT(U,V)]\,,\\
     I[\cT(U,V)]&\equiv\frac{32}{\pi}  \int dR\, d\theta\, R^3 \sin^2 \theta \, (U^{-1}+U^{-2}V+U^{-2})\bar{D}_{1,1,1,1}(U,V) \cT(U, V) \bigg|_{\substack{U = 1 + R^2 - 2 R \cos \theta \\
    V = R^2}}\,,
 }
 where 
 \es{D1111}{
 \bar{D}_{1,1,1,1}(U,V)=\frac{1}{z-\bar z}\left(\log(z \bar z)\log\frac{1-z}{1-\bar z}+2\text{Li}(z)-2\text{Li}(\bar z)\right)\,,
 }
 and $\mathcal{T}(U,V)$ is the reduced correlator for the 4-point function of the $\mathcal{N}=4$ stress tensor superprimary $S$:
 \es{2222}{
 & \langle S( x_1,Y_1) S( x_2,Y_2) S( x_3,Y_3) S( x_4,Y_4) \rangle = \frac{1}{\vec{x}_{12}^4 \vec{x}_{34}^{4}}
   \bigl[ 
  \Theta(U,V;Y)\cT(U,V)+\cS_{\text{free}}(U,V;Y)
    \bigr]\,.
    }
    Here, $Y^I$ are polarization vectors for $SU(4)_R\cong SO(6)_R$ vector indices $I=1,\dots 6$, the free theory contribution is $\cS_{\text{free}}$, and $\Theta(U,V;Y)$ is fixed by the superconformal Ward identities to be \cite{Dolan:2001tt}
 \es{redSText}{
 \Theta(U,V;Y) &\equiv  VY_{12}^2 +UVY_{13}^2 Y_{24}^2 + UY_{14}^2 Y_{23}^2 + U(U-V-1) Y_{13} Y_{14} Y_{23} Y_{24} \\
 &\quad +(1-U-V)Y_{12} Y_{14} Y_{23} Y_{34}  +V(V-U-1)Y_{12} Y_{13} Y_{24} Y_{34}\,.
}
To reduce to $\mathcal{N}=2$, we write the $SU(4)_R$ polarizations as $SU(2)_R\times SU(2)_F$ polarizations by setting
\begin{align}
Y^I = \frac{1}{\sqrt{2}}y^\alpha\bar{y}^{\dot{\alpha}} \sigma^I_{\alpha\dot{\alpha}}\ecq Y^5=Y^6=0 \ec \label{Y2yy}
\end{align}
where $\sigma^I_{\alpha\dot{\alpha}}$ for $I=1,\ldots,4$ are defined in terms of the usual Pauli matrices as $\sigma^I_{\alpha\dot{\alpha}}\equiv(1,i\sigma^1,i\sigma^2,i\sigma^3)$, and we introduced the $SU(2)_F$ polarizations $\bar{y}^{\dot{\alpha}}$ similar to the $SU(2)_R$ polarizations ${y}^{{\alpha}}$ considered above. It is easy to verify that the ansatz \eqref{Y2yy} respects the condition $Y\cdot Y=0$ that the ${SO}(6)_R$ polarizations $Y^I$ must satisfy. Upon performing this decomposition and comparing to the $\mathcal{N}=2$ reduced correlator $\mathcal{G}(U,V)$ in \eqref{ward4dN2}, we find that
\es{N4toN2}{
\mathcal{G}^\text{int}(U,V,\bar y)&= \mathcal{T}(U,V)\Big(\frac{(1-U+V)}{2} \langle\bar y_1,\bar y_2\rangle^2\langle\bar y_3,\bar y_4\rangle^2\\
&\quad+ \frac{(-1+U+V)}{2}  \langle\bar y_1,\bar y_3\rangle^2\langle\bar y_2,\bar y_4\rangle^2+ \frac{(1+U-V)}{2}  \langle\bar y_1,\bar y_4\rangle^2\langle\bar y_2,\bar y_3\rangle^2 \Big)\,,
}
where we wrote the flavor dependence in terms of the $SU(2)_F$ polarizations. After summing over these flavor indices and using the $\mathcal{N}=4$ relation $c=k/4$ we find that 
\es{sumoverP}{
k^2\sum_r P_r^{1111} \mathcal{G}^\text{int}_r(U,V)=32c^2\mathcal{T}(U,V)(1+U+V)\,,
}
where the projectors for $SU(2)_F$ are normalized as \eqref{projNorm} and the index $1$ refers to the $SU(2)_F$ mass $m$ whose derivatives we take. We can plug this relation into \eqref{simple} and compare to \eqref{constraint2} to fix $ {\bf f}(U,V)$ and thereby derive the general $\mathcal{N}=2$ integrated constraint
 \es{d4FCombinedAgain}{
 &-\partial_{m_{A}} \partial_{m_{B}} \partial_{m_{C}} \partial_{m_{D}} F\big|_{m=0} =-\partial_{m_{A}} \partial_{m_{B}} \partial_{m_{C}} \partial_{m_{D}} F\big|^\text{free}_{m=0}+\sum_r P_r^{(ABCD)}I[\mathcal{G}^\text{int}_r(U,V)]\,,\\
&\qquad\qquad\qquad  I[\mathcal{G}^\text{int}_r]\equiv\frac{k^2}{ \pi}  \int dR\, d\theta\, R^3 \sin^2 \theta
     \frac{\bar D_{1,1,1,1}(U,V)\cG^\text{int}_r(U, V)}{U^2} 
\bigg|_{\substack{U = 1 + R^2 - 2 R \cos \theta \\
    V = R^2}}\,.
 }
Note that not all choices of mass derivatives are independent. The number of independent constraints is given by the number of quartic casimirs. For instance, $SU(2)$ has just one quartic casimir, which is why there was just a unique integrated constraint of this type for $\mathcal{N}=4$ SYM.\footnote{A second integrated constraint on the stress tensor multiplet 4-point for $\mathcal{N}=4$ can be derived by also taking derivatives of the complex coupling $\tau$ \cite{Binder:2019jwn}, but for general $\mathcal{N}=2$ CFTs $\tau$ couples to the chiral multiplet, which is not related to the flavor current multiplet.} For any flavor group with rank greater than one, which includes all known $\mathcal{N}=2$ gauge theories other than $\mathcal{N}=4$ SYM, there are at least two quartic casimirs, so there will be at least two independent integrated constraints.

\subsection{SQCD }
\label{intConSQCD}

We will now consider the case of $SU(2)$ SQCD with $G=SO(8)$, for which we will give explicit formulae for both sides of the constraint \eqref{d4FCombinedAgain}. Without loss of generality, we can restrict to masses corresponding to the 4 Cartans:
\es{massMat}{
A=1,2,3,4:\qquad m_A= T^{ab}_A\mu_A\,,\qquad T^{ab}_A=-i(\delta_{a,2A-1}\delta_{b,2A}-\delta_{b,2A-1}\delta_{a,2A})\,,
}
where $a,b=1,\dots8$ are fundamental indices. The three quartic Casimirs for $SO(8)$ can then be computed as
\es{quartCas}{
\frac12\tr({m}^4)&=\sum_{A=1}^4 \mu_A^4\,,\\
 \frac14(\tr(m^2))^2-\frac12\tr(m^4)&=\sum_{A\neq B=1}^4\mu^2_A\mu^2_B\,,\\
  \frac{1}{2^44!}\varepsilon_{a_1,\dots,a_8}m_{a_1,a_2}m_{a_3,a_4}m_{a_5,a_6}m_{a_7,a_8}&=\mu_1\mu_2\mu_3\mu_4\,,
}
where $m\equiv \sum_{A=1}^4 T^{ab}_A\mu_A$ is a matrix with fundamental indices that we trace over.
We thus have three integrated constraints:
\es{SQCDints}{
  -\partial_{\mu_1}^4 F\big|_{\mu=0}
   &= 24\zeta(3)+\sum_r P_r^{1111}I[\mathcal{G}^\text{int}_r(U,V)]
 \,,\\
   -\partial_{\mu_1}^2\partial_{\mu_2}^2 F\big|_{\mu=0}
   &= \sum_r P_r^{(1122)}I[\mathcal{G}^\text{int}_r(U,V)]
 \,,\\
    -\partial_{\mu_1}\partial_{\mu_2}\partial_{\mu_3}\partial_{\mu_4} F \big|_{\mu=0}
   &= \sum_r P_r^{(1234)}I[\mathcal{G}^\text{int}_r(U,V)]
 \,,\\
 }
where we note that the free theory contribution is only nonzero for the first term. We will find it convenient to consider the linear combinations
\es{Fcsv}{
\mathcal{F}_{\bf v}&\equiv   -4\partial_{\mu_1}^2\partial_{\mu_2}^2 F\big|_{\mu=0}\,,\\
 \mathcal{F}_{\bf c}&\equiv    -\partial_{\mu_1}^4 F\big|_{\mu=0}- \partial_{\mu_1}^2\partial_{\mu_2}^2 F\big|_{\mu=0}+2  \partial_{\mu_1}\partial_{\mu_2}\partial_{\mu_3}\partial_{\mu_4} F \big|_{\mu=0}\,,\\
  \mathcal{F}_{\bf s}&\equiv    -\partial_{\mu_1}^4 F\big|_{\mu=0}- \partial_{\mu_1}^2\partial_{\mu_2}^2 F\big|_{\mu=0}-2  \partial_{\mu_1}\partial_{\mu_2}\partial_{\mu_3}\partial_{\mu_4} F \big|_{\mu=0}\,,\\
}
since after evaluating the sum over projectors using \eqref{SO8Proj} we get
\es{Fcsv2}{
\mathcal{F}_{\bf v}&=\sum_r I_{\bf v}[ \cG^\text{int}_r ]\,,\quad \mathcal{F}_{\bf c}=24\zeta(3)+\sum_r I_{\bf c}[ \cG^\text{int}_r ]\,,\quad \mathcal{F}_{\bf s}=24\zeta(3)+\sum_r I_{\bf s}[ \cG^\text{int}_r ]\,,\\
 I_{\bf v}[\vec g]&\equiv \begin{pmatrix} \frac{1}{21} I[g_{\bf 1}] & 0 & \frac{2}{9} I[g_{\bf 35_c}]  & \frac{2}{9}  I[g_{\bf 35_s}] & -\frac{1}{9}  I[g_{\bf 35_v}] & 0 & \frac{20}{21}  I[g_{\bf 300}]  \end{pmatrix}\,, \\
  I_{\bf c}[\vec g]&\equiv \begin{pmatrix} \frac{1}{21} I[g_{\bf 1}] & 0 & -\frac{1}{9} I[g_{\bf 35_c}]  & \frac{2}{9}  I[g_{\bf 35_s}] & \frac{2}{9}  I[g_{\bf 35_v}] & 0 & \frac{20}{21}  I[g_{\bf 300}]  \end{pmatrix}\,, \\
 I_{\bf s}[\vec g]&\equiv \begin{pmatrix} \frac{1}{21} I[g_{\bf 1}] & 0 & \frac{2}{9} I[g_{\bf 35_c}]  & -\frac{1}{9}  I[g_{\bf 35_s}] & \frac{2}{9}  I[g_{\bf 35_v}] & 0 & \frac{20}{21}  I[g_{\bf 300}]  \end{pmatrix}\,, \\
}
which are permuted by triality.\footnote{The $24\zeta(3)$ terms take into account the difference between the full correlator $\cG$ and the interacting part $\cG^\text{int}$, where the full correlator is what is permuted by triality.}

The left-hand sides of these integrated constraints are written in terms of the mass-deformed free energy $F(\mu_A, \tau, \bar \tau)$.  The partition function $Z(\mu_A,\tau,\bar\tau)\equiv \exp(-F(\mu_A,\tau,\bar\tau))$ was computed using supersymmetric localization in \cite{Pestun:2007rz,Alday:2009aq} for an $\mathcal{N}=2$ theory with gauge group $\mathfrak{G}$ in terms of a $\text{rank}(\mathfrak{G})$-dimensional integral. For $SU(2)$ SQCD we have a single integral\footnote{Our $\mu_I$ and $a_i$ are related to those in \cite{Alday:2009aq} by the redefinition $a_j\to -ia_j$ and $\mu\to i\mu+1$.}
     \es{MatrixModelSQCDSO8}{
	Z(m, \tau,\bar\tau)
	&= \int d^2 a\, \delta(a_1+a_2) \frac{ \prod_{i < j}(a_i  - a_j)^2 H^2(a_i - a_j)}{\prod_i\prod_{A=1}^{4} H(a_i + \mu_{A}) } 
	e^{-\frac{8 \pi^2  }{g_\text{UV}^2} \sum_i a_i^2} 
	\abs{Z_\text{inst}(m,a,\tau_\text{UV})}^2 \,, 
	}
	where $H(m) \equiv e^{-(1 + \gamma) m^2} G(1 + im) G(1 - im)$ is written in terms of the Barnes G-function $G(z)$ and the Euler-Mascheroni constant $\gamma$, and the $SU(2)$ gauge group has indices $i,j=1,2$. The term $Z_\text{inst}(m,a,\tau)$ encodes the contribution from instantons localized at the north pole of $S^4$. 
	
	There are two subtleties in evaluating this expression. Firstly, the complex coupling $\tau_\text{UV}\equiv i\frac{4\pi}{g_\text{UV}^2}+\frac{\theta_\text{UV}}{2\pi}$ that appears in the UV lagrangian does not transform simply under the $SL(2,\mathbb{Z})$ duality group of the IR CFT. Instead, the parameter that transforms in the usual way under $SL(2,\mathbb{Z})$ is the IR coupling $\tau_\text{IR}$ derived from the Seiberg-Witten curve as \cite{Grimm:2007tm}
   \es{IRUV}{
   e^{i\pi\tau_\text{IR}}=e^{-\frac{\pi  K\left(1-e^{2 i \pi  \tau_\text{UV}}\right)}{K\left(e^{2 i \pi  \tau_\text{UV}}\right)}}\,,\qquad \tau_\text{UV}=\frac{2 \log ( \frac{\theta_2(e^{i\pi\tau_\text{IR}})}{\theta_3(e^{i\pi\tau_\text{IR}})}  )}{ \pi i}\,,
   }
where $K(z)$, $\theta_2(z)$, and $\theta_3(z)$ are elliptic functions. Since we are mostly interested in $\tau_\text{IR}$, we will drop its subscript in what follows. 

 \begin{figure}[]
\begin{center}
       \includegraphics[width=.49\textwidth]{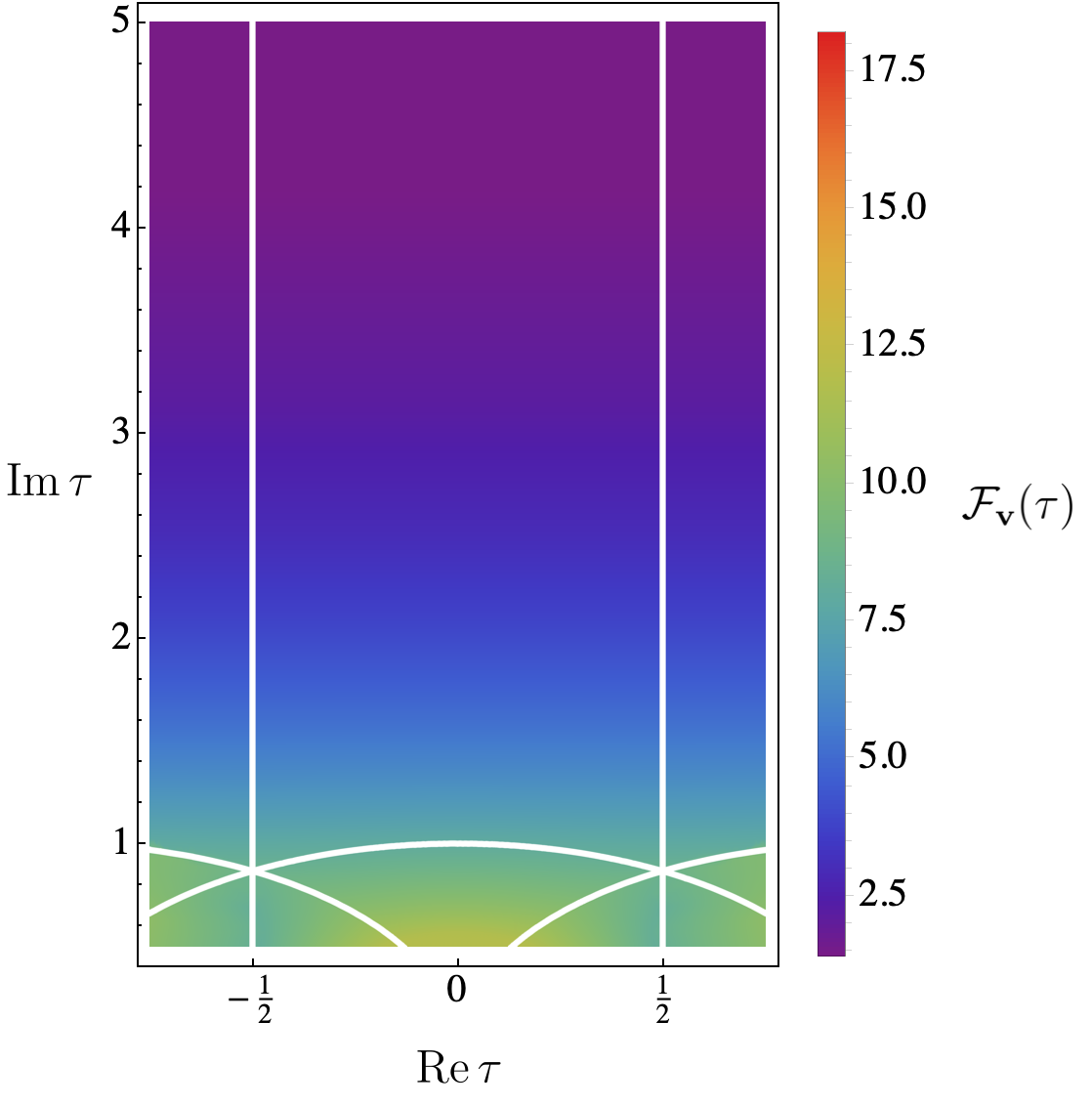}
      \includegraphics[width=.49\textwidth]{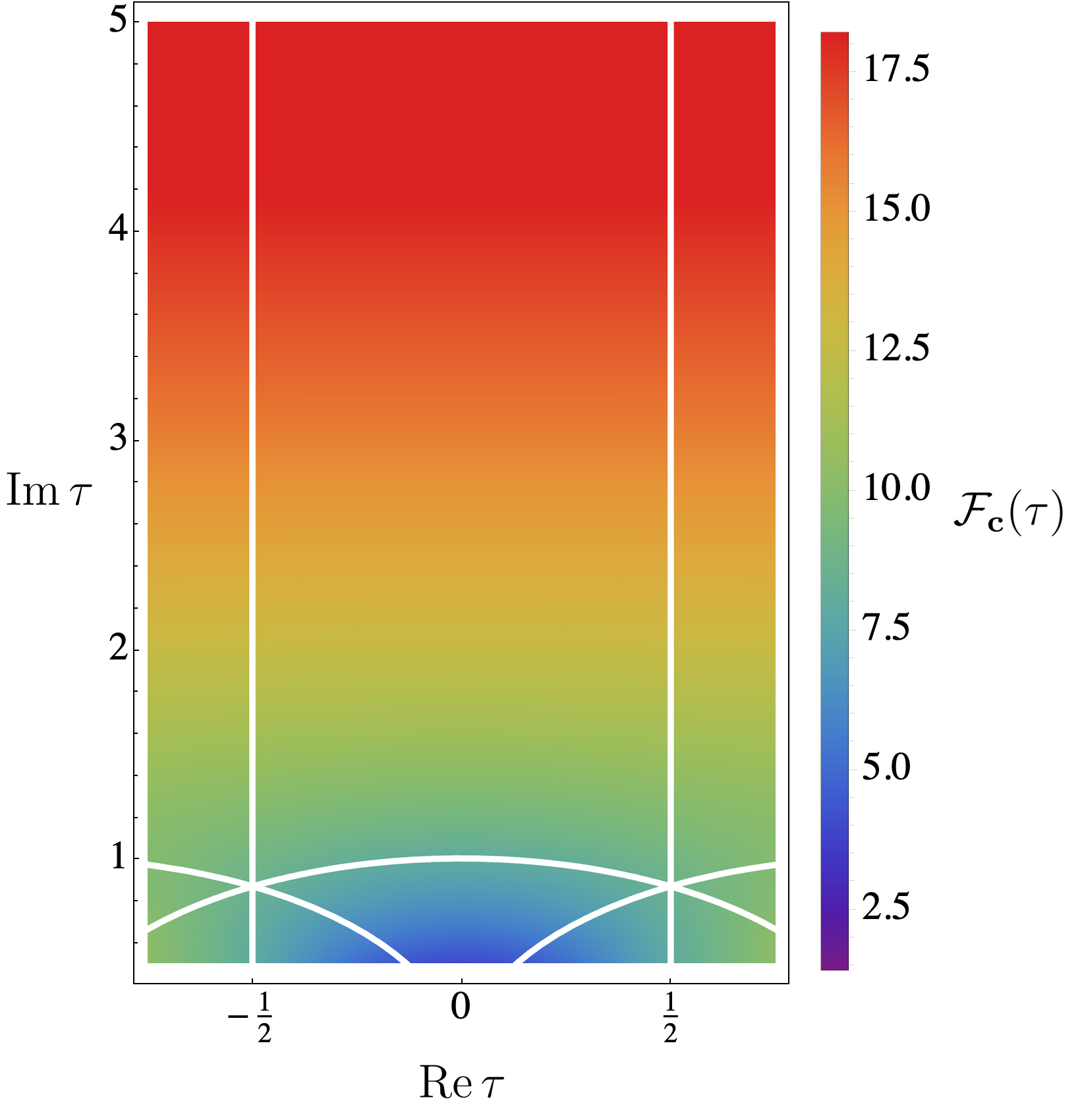}
            \includegraphics[width=.49\textwidth]{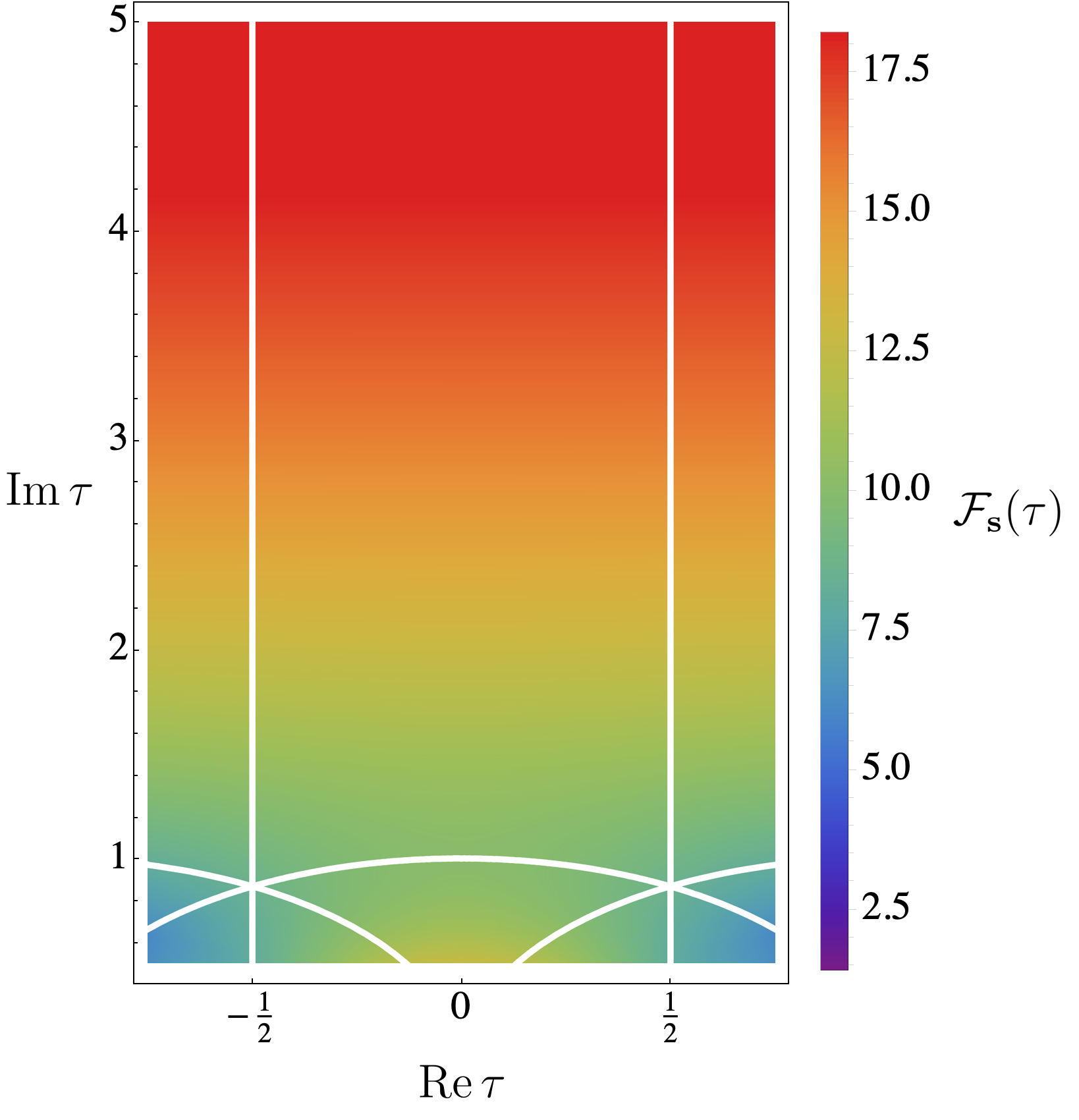}
\caption{The localization inputs $\mathcal{F}_{\bf v}$, $\mathcal{F}_{\bf c}$, and $\mathcal{F}_{\bf s}$ as a function of the complex coupling $\tau$. White lines show boundaries between some of the fundamental domains of $SL(2,\mathbb{Z})$, with a focus on the standard fundamental domain $|\tau| > 1$, $|\Re \tau| \leq \frac{1}{2}$. Note that the triality related localization inputs in different duality fundamental domains are related as in \eqref{dualTrial}.}
\label{fundF}
\end{center}
\end{figure}  

 \begin{figure}[]
\begin{center}
       \includegraphics[width=.49\textwidth]{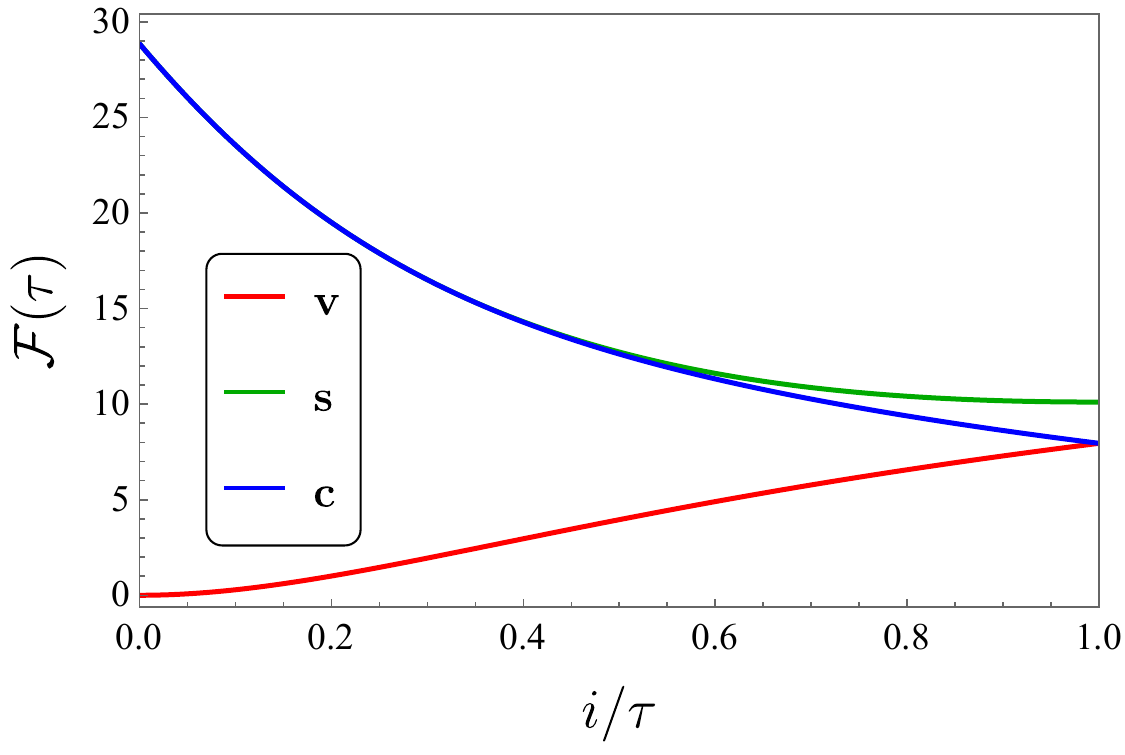}
      \includegraphics[width=.49\textwidth]{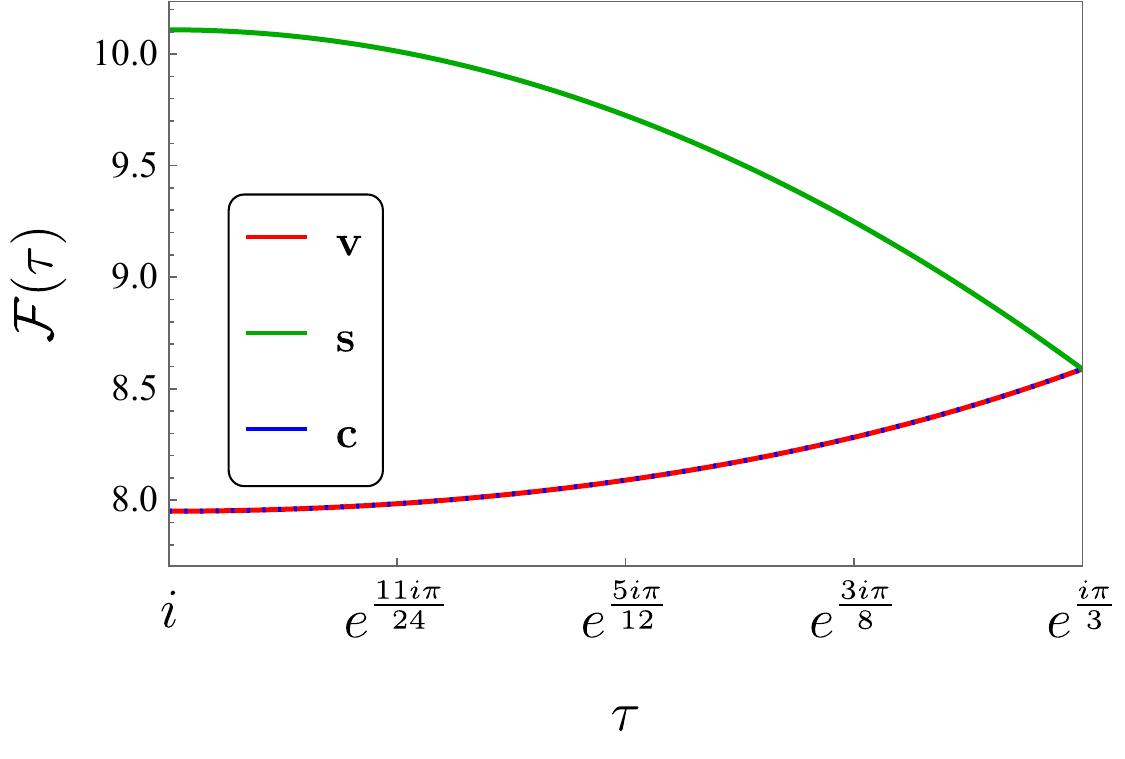}
\caption{The localization inputs for $\mathcal{F}_{\bf v}$, $\mathcal{F}_{\bf c}$, and $\mathcal{F}_{\bf s}$ as a function of $i/\tau $ from the free point $\tau=i\infty$ until the $S$-duality invariant point $\tau=i$ with $\Re(\tau)=0$ (\textbf{left}), and as a function of $\tau$ along the $S$-duality invariant arc with $|\tau|=1$ from $\tau=i$ to the $T$-duality invariant point $\tau=e^{\pi i/3}$ (\textbf{right}). Note that the triality related localization inputs in different duality fundamental domains are related as in \eqref{dualTrial}.}
\label{arcF}
\end{center}
\end{figure}  

The second subtlety in computing the localization expression is that $Z_\text{inst}(m,a,\tau)$ has been computed for $U(N)$ gauge groups \cite{Nekrasov:2002qd,Nekrasov:2003rj}, but not $SU(N)$. This was resolved by AGT in \cite{Alday:2009aq}, where it was shown that the partition function of $SU(2)$ SQCD can be computed in terms of a correlation function of 2d Liouville theory to get
     \es{MatrixModelSQCDSO82}{
	Z(m, \tau,\bar\tau)
	&= e^{(\text{quadratic in $m_A$})}\int d a \frac{ a^2(16)^{2a^2} H^2(2a)e^{-2\pi\text{Im}(\tau) a^2} }{ \prod_{A=1}^{4} H(a + \mu_{A})H(-a + \mu_{A}) } 	
	\abs{{\bf H}(m,a,\tau)}^2 \,, 
	}
	where ${\bf H}(m,a,\tau)$ can be computed as an expansion in $e^{i\pi\tau}$ using Zamolodchikov's recursion relation for Virasoro blocks \cite{Zamolodchikov:1984eqp}, as we review in Appendix \ref{locApp}. The discrepancy between the instanton factor for $U(2)$ versus $SU(2)$ ends up contributing only to the $a$-independent prefactor, which vanishes when we take four derivatives of mass. Note that this expression is naturally written in terms of the IR coupling $\tau$, and the expansion for ${\bf H}(m,a,\tau)$ converges rapidly in the $SL(2,\mathbb{Z})$ fundamental domain:
\es{tauDomain}{
|\tau|\geq1\,,\qquad |\Re(\tau)|\leq\frac12\,.
}
We plot the localization inputs $\mathcal{F}_{\bf v}$, $\mathcal{F}_{\bf c}$, and $\mathcal{F}_{\bf s}$ as a function of $\tau$ within this fundamental domain in Figure \ref{fundF}. Figure \ref{arcF} additionally show cross sections of these inputs along the imaginary axis from the free theory point $\tau=i\infty$ until the $S$-duality invariant point $\tau=i$, and on the arc from $\tau=i$ to the $S$ and $T$ duality invariant point $\tau=e^{\pi i/3}$. As shown in \cite{Seiberg:1994aj}, the $SL(2,\mathbb{Z})$ duality acts on the $SO(8)$ triality frame (and thus the localization inputs) as
\es{dualTrial}{
&S:\qquad\tau\to-1/\tau\qquad \Leftrightarrow \qquad{\bf 35_v} \leftrightarrow {\bf 35_c} \qquad\Leftrightarrow   \qquad{\mathcal{F}_{\bf v}} \leftrightarrow {\mathcal{F}_{\bf c}}\,,\\
&T:\qquad\tau\to\tau+1\qquad \Leftrightarrow \qquad{\bf 35_c} \leftrightarrow {\bf 35_s}\qquad\Leftrightarrow   \qquad{\mathcal{F}_{\bf c}} \leftrightarrow {\mathcal{F}_{\bf s}}\,, \\
}
which we can see in the figures for the localization inputs. For instance, at the $S$-duality invariant point $\tau=i$, we see that $\mathcal{F}_{\bf v}$ equals $\mathcal{F}_{\bf c}$, which continues along the arc $|\tau|=1$ until the point $\tau=e^{\pi i/3}$ which is also $T$-duality invariant, and where we see that all three localization inputs meet.

\section{Bootstrapping SQCD with integrated constraints}
\label{intBoot}

We will now apply the integrated constraints derived in the previous section to numerically bootstrap $SU(2)$ SQCD. We start by deriving the crossing equations, and show how they can combined with the integrated constraints to bound CFT data using the approach first developed for $\mathcal{N}=4$ SYM in \cite{Chester:2021aun}. We then show bounds for the scaling dimensions of the lowest dimension scalars in the ${\bf 35_c}$, ${\bf 35_v}$, and ${\bf 35_s}$ along the fundamental domain of $\tau$.

\subsection{Setup }
\label{setup}

The crossing equations from swapping the first and third operator in the moment map 4-point function \eqref{phiExp1} of a general CFT with flavor group $G$ were derived in \cite{Beem:2014zpa} and take the form 
\es{crossingSQCD}{
V^2 F_s^r\mathcal{G}_s(U,V)-U^2\mathcal{G}_r(V,U)+\frac{z\bar z}{z-\bar z}(z(\bar z-1)f_r(1-z)-\bar z(z-1)f_r(1-\bar z))=0\,,
}
where the crossing matrix $F_r^s$ is defined in terms of the projectors \eqref{flavorProj} as
\es{Fdef}{
P_r^{ABCD}=P_s^{CBAD}F_r^s\,,
}
and the holomorphic correlator $f(z)$ is given in \eqref{freeGen}. For $G=SO(8)$ the crossing matrix can be derived from the projectors in \eqref{SO8Proj} to get
\es{FSO8}{
F=
\left(
\begin{array}{ccccccc}
 \frac{1}{28} & \frac{1}{28} & \frac{1}{28} &
   \frac{1}{28} & \frac{1}{28} & \frac{1}{28}
   & \frac{1}{28} \\
 1 & \frac{1}{2} & \frac{1}{3} & \frac{1}{3}
   & \frac{1}{3} & 0 & -\frac{1}{6} \\
 \frac{5}{4} & \frac{5}{12} & \frac{7}{12} &
   -\frac{5}{12} & -\frac{5}{12} &
   -\frac{1}{12} & \frac{1}{12} \\
 \frac{5}{4} & \frac{5}{12} & -\frac{5}{12} &
   \frac{7}{12} & -\frac{5}{12} &
   -\frac{1}{12} & \frac{1}{12} \\
 \frac{5}{4} & \frac{5}{12} & -\frac{5}{12} &
   -\frac{5}{12} & \frac{7}{12} &
   -\frac{1}{12} & \frac{1}{12} \\
 \frac{25}{2} & 0 & -\frac{5}{6} &
   -\frac{5}{6} & -\frac{5}{6} & \frac{1}{2}
   & -\frac{1}{3} \\
 \frac{75}{7} & -\frac{25}{14} & \frac{5}{7}
   & \frac{5}{7} & \frac{5}{7} & -\frac{2}{7}
   & \frac{3}{14} \\
\end{array}
\right)
\,.
}
We can then plug the block expansion \eqref{blockExpSQCD} of $\mathcal{G}_r$ into \eqref{crossingSQCD} to derive $i=1,\dots 7$ crossing equations:
\es{crossingSQCD2}{
0= V^i_\text{short}(U,V)+\sum_{\ell}\sum_{\Delta\geq\ell+2}\sum_{r}\lambda^2_{\Delta,\ell,r} V^i_r[U^{-1}G_{\Delta+2,\ell}(U,V)]\,,
}
where $\ell$ is even for ${\bf 1}$, ${\bf 35_v}$, ${\bf 35_c}$, ${\bf 35_s}$ and ${\bf 300}$, and odd for ${\bf 28}$ and ${\bf 350}$, and the crossing functions are
\es{Vs}{
\vec V_{{\bf1}}[g]&=\begin{pmatrix} \frac{29}{28} F_-[g]&\frac{1}{28}F_-[g]&\frac{1}{28}F_-[g]&\frac{1}{28}F_-[g]&\frac{1}{28}F_-[g]&-\frac{27}{28}F_+[g]&\frac{1}{28}F_+[g]\end{pmatrix}\,,\\
\vec V_{{\bf28}}[g]&=\begin{pmatrix} F_-[g]&\frac{3}{2} F_-[g] &\frac{1}{3} F_-[g] &\frac{1}{3}F_-[g]&\frac{1}{3}F_-[g]& F_+[g]&-\frac{1}{2}F_+[g]\end{pmatrix}\,,\\
\vec V_{{\bf35_c}}[g]&=\begin{pmatrix}\frac{5}{4} F_-[g] &\frac{5}{12} F_-[g]  &\frac{19}{12} F_-[g] &-\frac{5}{12}  F_-[g]  &-\frac{5}{12} F_-[g] &\frac{5}{4} F_+[g] &\frac{5}{  12} F_+[g] \end{pmatrix}\,,\\
\vec V_{{\bf35_s}}[g]&=\begin{pmatrix}\frac{5}{4} F_-[g]  &\frac{5}{12} F_-[g]  &-\frac{5}{12}F_-[g]  &\frac{19}{12} F_-[g]  &  -\frac{5}{12}F_-[g]  &\frac{5}{4} F_+[g] &\frac{5}{  12} F_+[g] \end{pmatrix}\,,\\
\vec V_{{\bf35_v}}[g]&=\begin{pmatrix}\frac{5}{4}F_-[g]  &\frac{5}{12} F_-[g] &-\frac{5}{12}F_-[g] &-\frac{5}{12} F_-[g] &\frac{19}{12}F_-[g]  &\frac{5}{4} F_+[g] &\frac{5}{12} F_+[g] \end{pmatrix}\,,\\
\vec V_{{\bf350}}[g]&=\begin{pmatrix}\frac{25}{2}F_-[g]  &0&-\frac{5}{6} F_-[g]  &-\frac{5}{6} F_-[g]&-\frac{5}{6} F_-[g] &\frac{25}{2} F_+[g] &0\end{pmatrix}\,,\\
\vec V_{{\bf300}}[g]&=\begin{pmatrix}\frac{75}{7} F_-[g] &-\frac{25}{14} F_-[g] &\frac{5}{7} F_-[g]&\frac{5}{7} F_-[g] &\frac{5}{7}F_-[g]& \frac{75}{7} F_+[g] & -\frac{25}{14}F_+[g]\end{pmatrix}\,,\\
}
and for a given function $g$ we define
\es{F3}{
F_\pm[g(U,V)]&\equiv UV^{3}g(U,V)\pm VU^{3}g(V,U)\,.
}
 The short piece $\vec V_\text{short}(U,V)$ is a vector of exact functions given by applying crossing to the resummed short terms \eqref{shortsSO8} in $\cG_r$ as well as the last term in \eqref{crossingSQCD}. The expression for $\vec V_\text{short}(U,V)$ is rather messy, so we give it in the attached \texttt{Mathematica} file.

We will now impose both the integrated constraints and the crossing equations by reformulating them as an optimization problem following the numerical bootstrap approach of \cite{Chester:2021aun}. 
For a given function $ g_r(U,V)$ in irrep $r$, we consider the infinite-dimensional vector ${\bf x}_{g_r}$:
\es{Vinf}{
 {\bf x}_{g_r} \equiv \begin{pmatrix}   I_{\bf v} [g_r] \\
 I_{\bf c} [g_r] \\
  I_{\bf s} [g_r]  \\
  V^i_r [g(U,V)]
  \end{pmatrix}  \in {\bf X}_\infty\,.
}
The first three components of ${\bf x}_{g_r}$ are the functionals defined in \eqref{Fcsv2} and \eqref{d4FCombinedAgain} that appear in the integrated correlators.  The final entry denotes an infinite number of components labeled by $i=1,\dots,7$ and the continuous variables $U,V$. We can denote by ${\bf X}_\infty$ the vector space consisting of the vectors ${\bf x}_{g_r}$ of the form \eqref{Vinf}.

 The constraints \eqref{Fcsv2} then impose that an infinite sum of vectors with positive coefficients $\lambda^2_{\Delta,\ell,r}$ plus the contribution from the protected terms must vanish:
 \es{SumRule}{
 \sum_{\Delta,\ell,r}\lambda^2_{\Delta,\ell,r} {\bf x}_{\Delta, \ell,r} + {\bf w} = 0 \,, \qquad
  {\bf w} \equiv 
  \begin{pmatrix}
    \sum_r I_{\bf v}\left[\cF^r_\text{short}-{\mathcal{G}}^r_\text{free}\right]-\cF_{\bf v}(\tau,\bar\tau)\\
   \sum_r I_{\bf c}\left[\cF^r_\text{short}-{\mathcal{G}}^r_\text{free}\right]+24\zeta(3)-\cF_{\bf c}(\tau,\bar\tau)\\
     \sum_r I_{\bf s}\left[\cF^r_\text{short}-{\mathcal{G}}^r_\text{free}\right]+24\zeta(3)-\cF_{\bf s}(\tau,\bar\tau)\\
   V^i_\text{short}(U,V)
  \end{pmatrix}   \,,
}
where ${\bf x}_{\Delta, \ell,r}$ is a shorthand notation for ${\bf x}_{g_r}$ with $g_r(U, V)=U^{-1}G_{\Delta+2, \ell}(U, V)$, and we subtract ${\mathcal{G}}^r_\text{free}$ as given in \eqref{freeGSO8} because the integrated constraints were defined to act on ${\mathcal{G}}_\text{int}$.  To find instances in which the equation \eqref{SumRule} cannot be obeyed, consider a functional (written as an infinite-dimensional row vector)
 \es{alphaDef}{
   \alpha=\begin{pmatrix} \alpha_{\bf v} &  \alpha_{\bf c}  &  \alpha_{\bf s}  & \alpha_\infty \end{pmatrix}  \,,
  } 
where $ \alpha_{\bf v} $, $ \alpha_{\bf c} $, and  $\alpha_{\bf s}$  are numbers, and $\alpha_\infty$ is a functional acting on functions of $(U, V)$.  

The functional $\alpha$ can be used to bound scaling dimensions $\Delta$ by the following algorithm:
\\
\\
{\bf Scaling dimension bound:}
\begin{enumerate}
\item Normalize $\alpha$ such that $\alpha [{\bf w}] =1$.\footnote{The normalization $\alpha[ {\bf w}] =1$ is a matter of convention.  One can more generally fix $\alpha [{\bf w}]$ to any positive real number.} 
\item Assume that the scaling dimensions $\Delta_{\ell,r}$ of all spin $\ell$ unprotected operators in irrep $r$ obey lower bounds $\Delta_{\ell,r}\geq\bar\Delta_{\ell,r}$, where the unitarity bound $\Delta_{\ell,r}\geq\ell+2$ provides a minimal choice.
\item Search for $\alpha$ obeying
\es{searchScal}{
 \alpha [{\bf x}_{\Delta, \ell,r}] \geq 0 
}
for all $\Delta \geq \bar \Delta_{\ell,r}$ and all $\ell$ and $r$.
\item If such $\alpha$ exists, then by positivity of $\lambda^2_{{\Delta,\ell,r}}$ we have
 \es{alphScal}{
 \sum_{\Delta,\ell,r}\lambda^2_{\Delta,\ell,r} \alpha[{\bf x}_{\Delta, \ell,r}] + \alpha[{\bf w}] > 0 \,,
}
which contradicts Eq.~\eqref{SumRule}, and so our assumptions $\Delta\geq\bar\Delta_{\ell,r}$ must be false. If we cannot find such an $\alpha$, then we conclude nothing.
\end{enumerate}
For instance, we can set the lower bounds $\bar\Delta_{\ell,r}$ to their unitarity values $\ell+2$ for all $\ell$ and $r$ except a certain $\ell',r'$, then by varying $\bar\Delta_{\ell',r'}$, $\tau$, and $\bar\tau$, this algorithm can be used to find an upper bound on $\bar\Delta_{\ell',r'}$, i.e.~on the scaling dimension of the lowest dimension operator with spin $\ell'$ in irrep $r'$, as a function of $\tau$ and $\bar\tau$. A similar algorithm can be used to bound OPE coefficients, but we will not use it in this paper.

To implement the above algorithms numerically, we must perform three truncations / discretizations. The first truncation is the number of spins, which we can achieve by imposing a simple cap $\ell_\text{max}$ on the range of spins we consider. The second truncation is on the space of functions of $(U,V)$ in the infinite dimensional vector space ${\bf X}_\infty$, which we replace with a finite-dimensional vector space ${\bf X}_\Lambda$ defined by replacing a general vector ${\bf x} \in {\bf X}_\infty$ with 
 \es{Vfinite}{
  {\bf x} = \begin{pmatrix}
   x_{\bf v} \\
   x_{\bf c} \\
      x_{\bf s} \\
   \partial_z^m \partial_{\bar z}^n \, x_i(U, V)\vert_{z=\bar z=\frac12} 
  \end{pmatrix}\,,
 }
where $m+n\leq\Lambda$, $m\leq n$, and $m+n$ odd for $i\leq5$ and even for $i=6,7$, since $F_\pm[G_{\Delta+2,\ell}(U,V)]$ in \eqref{F3} is symmetric/antisymmetric in $z\leftrightarrow \bar z$ and odd/even under crossing for $+/-$, respectively.\footnote{We can efficiently compute $\partial_z^m\partial_{\bar z}^nG_{\Delta,\ell}(U,V)\vert_{z=\bar z=\frac12}$ using the \texttt{scalar\_blocks} code, available online from the bootstrap collaboration. Note that their blocks differ from our conventions as $G^\text{ours}_{\Delta,\ell}(U,V)=(\ell+1)G^\text{theirs}_{\Delta,\ell}(U,V)$.} For sufficiently high $\ell_\text{max}$, the numerical bounds will monotonically improve with increasing $\Lambda$, which is what makes the numerical bootstrap rigorous.  Lastly, we make the continuum of $\Delta_{\ell,r}\geq\bar\Delta_{\ell,r}$ finite by imposing a cap $\Delta\leq\Delta_\text{max}$ and discretizing $\Delta$ with spacing $\Delta_\text{sp}$. For the bounds in this paper, we used $\Lambda=27$, $\ell_\text{max}=30$, $\Delta_\text{max}=40$, and $\Delta_\text{sp}=.01$. We then solved the resulting set of linear constraints using \texttt{SDPB} \cite{Simmons-Duffin:2015qma,Landry:2019qug} as discussed in \cite{Chester:2021aun}, with the same \texttt{SDPB} parameters discussed in that paper.

For the integrated constraints, one needs to restrict the integration region in $I$ as defined in \eqref{d4FCombinedAgain} to a domain where the block expansion converges. Suitable domains were discussed in detail in \cite{Chester:2021aun}, to which we refer the reader for more details. We used the non-oscillating region $D'$ defined in that paper, for which we also checked that we obtained positive functionals for $\Delta_{\ell,r}\geq\bar\Delta_{\ell,r}$ as required for rigorous bounds. 

\subsection{Results }
\label{results}

 \begin{figure}[]
\begin{center}
       \includegraphics[width=.8\textwidth]{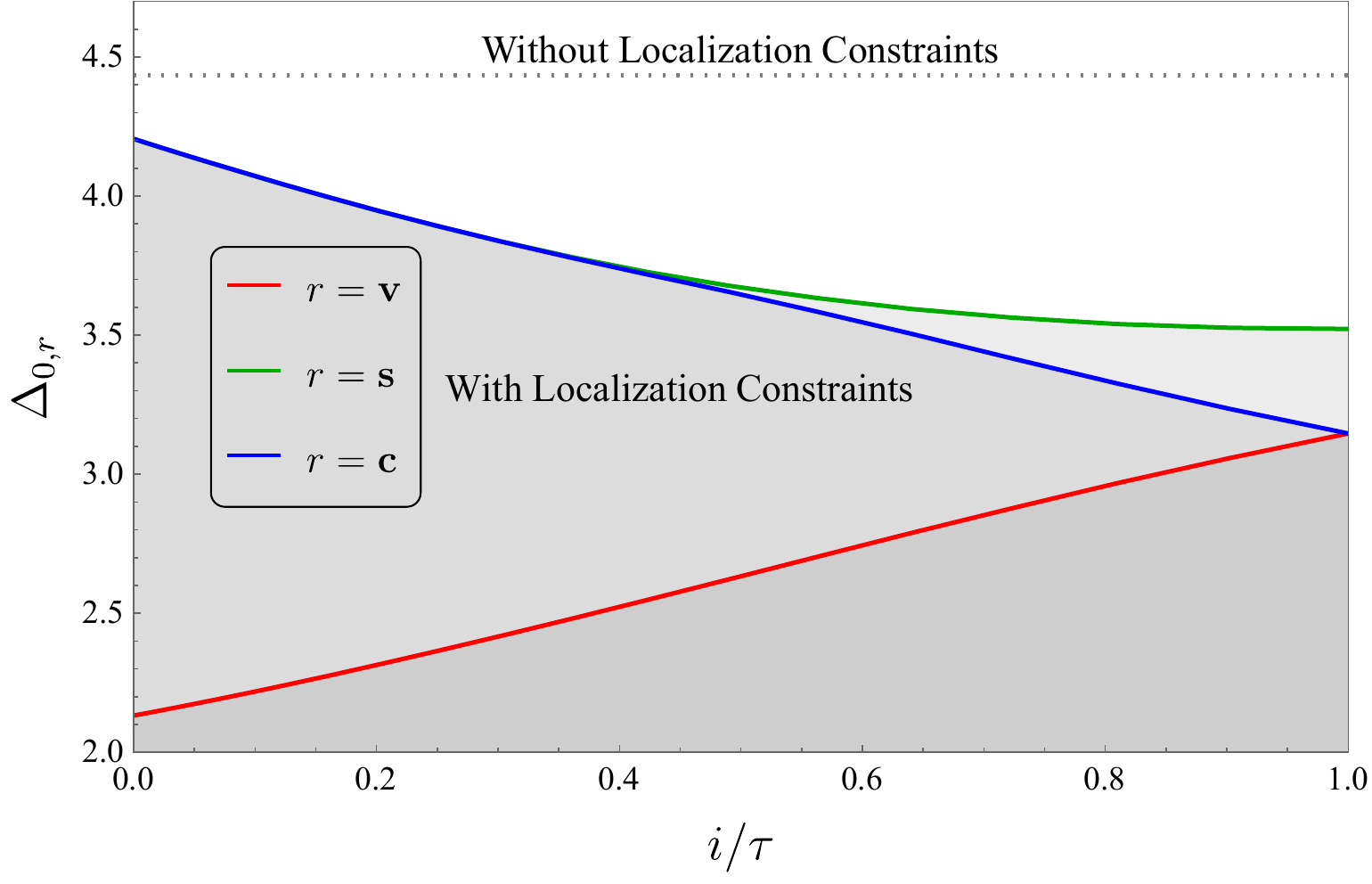}
       \\
       \vspace{.3in}
      \includegraphics[width=.8\textwidth]{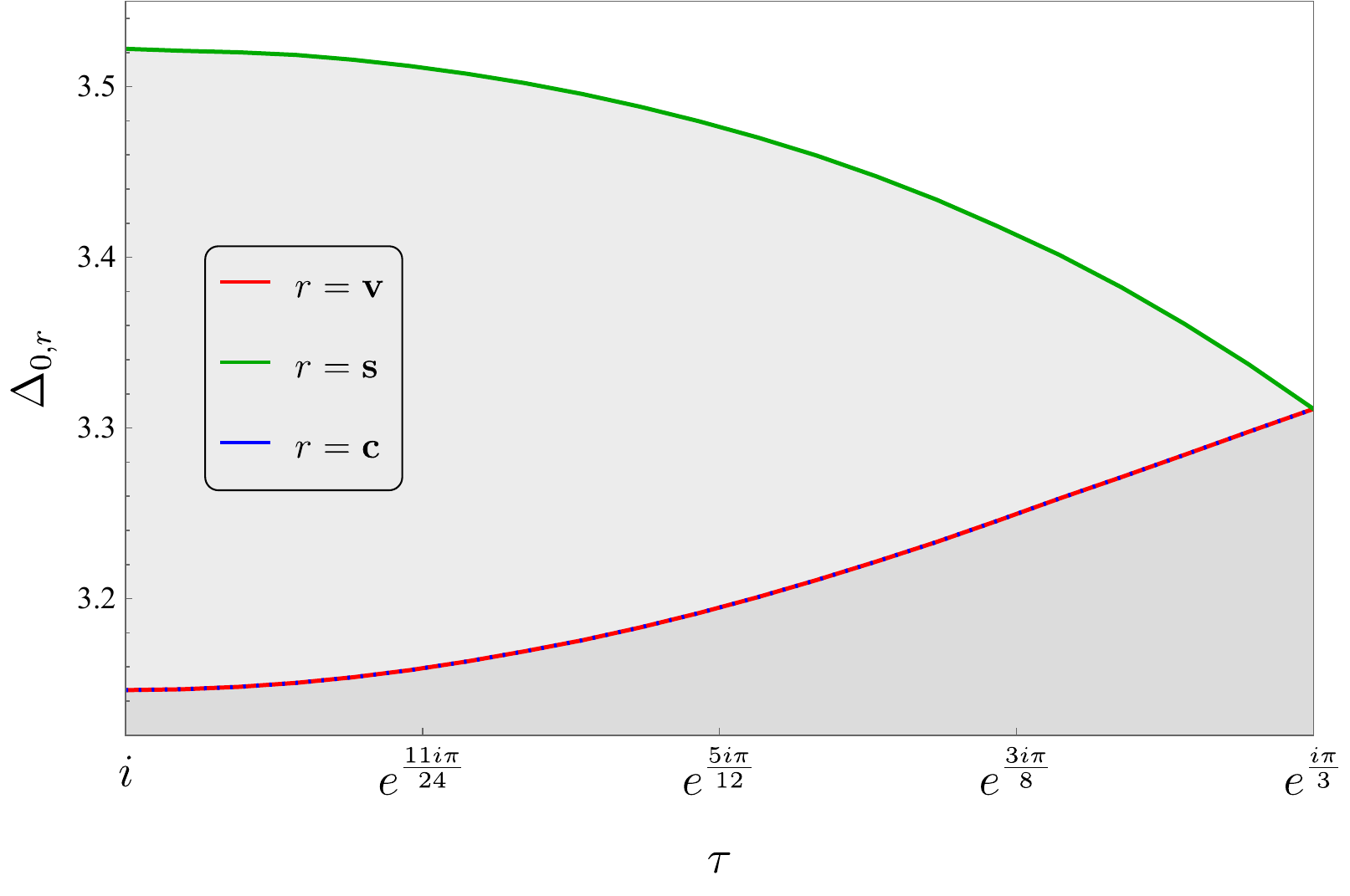}
	\caption{Upper bounds on the scaling dimension $\Delta_{0,r}$ of the lowest scalar in irreps $r={\bf 35_c},{\bf 35_s},{\bf 35_v}$ as a function of $i/\tau $ from the free point $\tau=i\infty$ until the $S$-duality invariant point $\tau=i$ with $\Re(\tau)=0$ (\textbf{top}), and as a function of $\tau$ along the $S$-duality invariant arc with $|\tau|=1$ from $\tau=i$ to the $T$-duality invariant point $\tau=e^{\pi i/3}$ (\textbf{bottom}). These bounds were computed with bootstrap resolution $\Lambda=27$, as defined below \eqref{Vfinite}. Note that the triality related bounds in different duality fundamental domains are related as in \eqref{dualTrial}. We also show the upper bounds that would be obtained at the same bootstrap resolution without using the integrated constraints, which are the same for ${\bf 35_c}$, ${\bf 35_v}$, and ${\bf 35_s}$.}
\label{dels}
\end{center}
\end{figure}  

Using the method outlined above, we can obtain upper bounds on the scaling dimensions $\Delta_{0,r}$ of the lowest scalar as a function of complex coupling $\tau$ for the irreps $r={\bf1}\,, {\bf 300}\,, {\bf 35_c}\,,{\bf 35_v}\,,{\bf 35_s}$, which are the irreps in the symmetric product of the adjoint ${\bf 28}$. 

We plot the three triality related 35-dimensional irreps in Figure \ref{dels}, where we focus on the slice of the fundamental domain from the free theory point $\tau=i\infty$ until the $S$-duality invariant point $\tau=i$ with $\Re(\tau)=0$, and then along the $S$-duality invariant arc with $|\tau|=1$ from $\tau=i$ to the $T$-duality invariant point $\tau=e^{\pi i/3}$. Recall that the three irreps are permuted by duality transformations as in \eqref{dualTrial}. For instance, for ${\bf 35_v}$ and $\Re(\tau)=0$, the upper bound is give by the red curve from $\tau=i\infty$ to $\tau=i$, and then by the blue curve from $\tau=i$ to $\tau=0$, since ${\bf 35_v}$ and ${\bf 35_c}$ are exchanged by the $S$-duality transformation $\tau\to-1/\tau$. Since each point is very computationally expensive,\footnote{On a node with 20 cores, \texttt{SDPB} takes 7 hours to run a singled feasible point, and these bounds were computed by running binary searches with 13 iterations.} we limited ourselves to these sections of the full fundamental domain, though from the localization input plots in Figure \ref{fundF}, we do not expect the $\Re(\tau)$ dependence to be significant for most of the range of $\Im(\tau)$. Note that the bounds with localization constraints do not meet the bounds without localization constraints for any value of $\tau$, similar to the $\mathcal{N}=4$ SYM case discussed in \cite{Chester:2021aun}.

We can test the convergence of these bounds by looking at the free point $\tau=i\infty$, where from the free correlator \eqref{freeGSO8} we expect $\Delta_{0,{\bf v}}=2$ and $\Delta_{0,{\bf c}}=\Delta_{0,{\bf s}}=4$, while with bootstrap resolution $\Lambda=27$ we get the bounds $\Delta_{0,{\bf v}}\leq2.13$ and $\Delta_{0,{\bf c}}=\Delta_{0,{\bf s}}\leq4.20$. The free theory thus differs from the boundary of the allowed region by a relative error of around $5\%$, which suggests that these bounds are not fully converged, but still reasonably accurate. The bounds for the $\Delta_{0,{\bf1}}$ and $\Delta_{0,{\bf300}}$ are much less converged, as we expect $2$ and $4$ from the free theory but find $2.5$ and $4.3$, respectively, so we leave further study of these irreps to a future project.

\section{Conclusion}
\label{conc}

The main result of this work is new exact relations between integrals of the moment map 4-point function in general 4d $\mathcal{N}=2$ CFTs and derivatives of the mass deformed sphere free energy $F(m)$, which for rank-$\mathfrak{G}$ gauge theories can be computed using supersymmetric localization in terms of a $\mathfrak{G}$-dimensional matrix model integral. The number of such independent integrated constraints was shown to be the number of quartic casimirs of the flavor group $G$, which is at least two for all known $\mathcal{N}=2$ gauge theories (except $\mathcal{N}=4$ SYM), and so is sufficient to fix the value of the complexified gauge coupling $\tau$ for simple gauge groups. These integrated constraints can thus be combined with the numerical conformal bootstrap to non-perturbatively bound CFT data in 4d $\mathcal{N}=2$ conformal simple gauge theories as a function of $\tau$, following the method introduced in \cite{Chester:2021aun}. We demonstrated this method for $SU(2)$ SQCD, which has $SO(8)$ flavor symmetry and thus three independent constraints. In particular, we computed non-perturbative bounds on the unprotected scaling dimensions $\Delta_{\bf 35_c}$, $\Delta_{\bf 35_s}$, and $\Delta_{\bf 35_v}$ as a function of $\tau$, which matched the free theory limit, and exhibited the expected mixing between the action of the $SL(2,\mathbb{Z})$ duality group and $SO(8)$ triality.

Looking ahead, one could improve the accuracy of the bounds by considering mixed correlators of the moment map operator with chiral operators of dimension $\Delta=p$ for $p=2,3,\dots$, whose bootstrap program was initiated in  \cite{Gimenez-Grau:2020jrx}. This mixed system will give us access to one nontrivial OPE coefficient per $p$ that can in principle be computed as a function of $\tau$ using localization \cite{Baggio:2014ioa,Baggio:2014sna,Baggio:2015vxa}, though the instanton contributions to the localization formula are only known so far for the $p=2$ case \cite{Gerchkovitz:2016gxx,Fucito:2015ofa}. For each $p$ we can also derive a new integrated constraint on the mixed corellator of moment map and chiral operators, which is the $\mathcal{N}=2$ version of the $\mathcal{N}=4$ constraints derived in \cite{Binder:2019jwn}. A feature of this mixed system is that protected operators appear that are isolated from the continuum of long operators, so one can bootstrap bootstrap both upper and lower bounds on their OPE coefficients, which could form islands as small as those found for ABJM theory in \cite{Agmon:2019imm}.

It would also be nice to check if our bootstrap bounds are approximately saturated by weak coupling results, which would give more evidence for the conjecture that these bounds are saturated by the physical theory in the infinite precision limit. For $\mathcal{N}=4$ SYM, the bootstrap bounds of \cite{Chester:2021aun} were indeed found to saturate the weak coupling results for the Konishi scaling dimension, which was computed to 4-loops in \cite{Velizhanin:2009gv}. So far, $SU(N)$ SQCD has only been studied in the planar (i.e. Veneziano) limit \cite{Gadde:2010zi}, but it should be straightforward to generalize these weak coupling calculations to finite $N$.

In this work, we only applied the general $\mathcal{N}=2$ bootstrap+localization formalism to $SU(2)$ SQCD. There is a huge zoo of 4d $\mathcal{N}=2$ gauge theories (see \cite{Argyres:2022mnu,Razamat:2022gpm} for recent reviews) to which this new method could be applied to. The most immediate generalization would be to $SU(N)$ SQCD, which has flavor group $U(2N)$, and thus four independent integrated constraints for $N>2$. For this one would need to derive the localization input for $N>2$, which can likely be done using the $SU(N)$ generalization of AGT as proposed in \cite{Wyllard:2009hg}. Numerical bootstrap bounds on $SU(N)$ SQCD could be essential to understanding the large $N$ limit of this theory at strong coupling, and further exploring possible holographic interpretations \cite{Gadde:2009dj}. 

Finally, we could also combine the integrated constraints derived in this work with the analytic bootstrap \cite{Rastelli:2017udc} to compute the large $N$ limit of the moment map 4-point function in holographic $\mathcal{N}=2$ CFTs, where the correrator is dual to gluon scattering. The leading large $N$ limit (and 1-loop correction), which is independent of $\tau$, was already considered in \cite{Zhou:2018ofp,Alday:2021ajh,Alday:2021odx,Behan:2022uqr}. To compute higher derivative corrections, which will depend on $\tau$, one can apply the integrated constraints given here, following similar work on graviton scattering from large $N$ $\mathcal{N}=4$ SYM correlators in \cite{Binder:2019jwn,Chester:2020dja,Chester:2019pvm,Chester:2019jas,Chester:2020vyz,Alday:2021vfb,Alday:2022uxp}.

\section*{Acknowledgments} 

We thank Ofer Aharony, Leonardo Rastelli, Silvu Pufu, Shlomo Razamat, Connor Behan, Alessandro Vichi and Ross Dempsey for useful conversations, Silviu Pufu and Ross Dempsey for collaboration at an early stage of this project, and Silviu Pufu, Ross Dempsey, and Ofer Aharony for reviewing the manuscript. SMC is supported by the Weizmann Senior Postdoctoral Fellowship. 

\appendix

\section{Supersymmetric localization}
\label{locApp}

In this appendix we discuss how to use the AGT relation \cite{Alday:2009aq} to compute the mass deformed partition function in \eqref{MatrixModelSQCDSO8}. AGT relates the instanton term  $Z_\text{inst}(m,a,\tau_\text{UV})$ to the holomorphic $c_\text{2d}=25$ Virasoro block $\mathcal{V}_{h}^{h_1,h_2,h_3,h_4}(q)$ as
 \es{AGTblock}{
   Z_\text{inst}(m,a,\tau_\text{UV})=(1-e^{2\pi i\tau_\text{UV}})^{2 \left(\sqrt{h_1-1}+1\right) \left(\sqrt{h_3-1}+1\right)}
   (e^{2\pi i\tau_\text{UV}})^{-h+{h_1}+{h_2}}\mathcal{V}_{h}^{h_1,h_2,h_3,h_4}(q)\,,
   }
   where $q\equiv e^{i\pi\tau}$ and we define
      \es{hs}{
   h&=1+a^2\,,\qquad h_1=1-\tilde m_0^2\,,\qquad h_2=(2-m_0)m_0\,,\qquad h_3=(2-m_1)m_1\,,\qquad h_4=1-\tilde m_1^2\,,\\
    i\mu_1+1&\equiv m_0+\tilde m_0\,,\qquad    i\mu_2+1\equiv m_0-\tilde m_0\,,\qquad    i\mu_3+1\equiv m_1+\tilde m_1\,,\qquad    i\mu_4+1\equiv m_1-\tilde m_1\,.
   }
The Virasoro block can be computed in a small $q$ expansion using Zamolodchikov's recursion relation as \cite{Zamolodchikov:1984eqp}
   \es{Zam}{
   \mathcal{V}_{h}^{h_1,h_2,h_3,h_4}(q)&= (16q)^{h-1}
   z^{1-h_1-h_2}
   (1-z)^{1-h_2-h_3} \theta
   _3(q){}^{12-4 (h_1+h_2+h_3+h_4)}{\bf H}(q)\,,\\
   }
   where here $z=e^{2\pi i \tau_\text{UV}}$. We can compute ${\bf H}(q)=\lim_{b\to1}{\bf H}(b,h,q)$ recursively as
   \es{}{
      {\bf H}(b,h,q)&=1 + \sum_{m,n\ge1} \frac{q^{mn} R_{m,n}}{h - h_{m,n}} {\bf H}(b, h_{m,n} + mn, q)\,,
      }
      where
      \begin{equation}
	 h_{i}=\frac{1}{4}\left(b+\frac{1}{b}\right)^2-\lambda_i^2,\quad h_{m,n}=\frac{1}{4}\left(b+\frac{1}{b}\right)^2-\lambda_{m,n}^2\,,\qquad \lambda_{m,n}=\frac{1}{2}\left(\frac{m}{b}+nb\right),
	\end{equation}
   and $R_{m,n}$ is given by 
\begin{equation}\label{eq:Rmn}
R_{m,n}=2\frac{\prod_{p,q}\left(\lambda_1+\lambda_2-\lambda_{p,q}\right)\left(\lambda_1-\lambda_2-\lambda_{p,q}\right)\left(\lambda_3+\lambda_4-\lambda_{p,q}\right)\left(\lambda_3-\lambda_4-\lambda_{p,q}\right)}{\prod_{k,l}'\lambda_{k,l}},
\end{equation}
and the ranges of $p,q,k,$ and $l$ are:
\begin{align*}
p&=-m+1,-m+3,\cdots,m-3,m-1,\\
q&=-n+1,-n+3,\cdots,n-3,n-1,\\
k&=-m+1,-m+2,\cdots,m,\\
l&=-n+1,-n+2,\cdots,n\,. 
\end{align*}
The prime on the product in the denominator means
that $\left(k,l\right)=\left(0,0\right)$ and $\left(k,l\right)=(m,n)$ are excluded. For instance, to the first nontrivial order in $q$ we find after using the relations \eqref{hs}:
\es{qex}{
{\bf H}(q) =1
+q\frac{8 \mu_1 \mu_2
   \mu_3 \mu_4 }{a^2+1}+O(q^2)\,,
}
and for the results shown in the main text we computed ${\bf H}(q) $ to $O(q^8)$, which allows us to compute the localization inputs to many digits of accuracy.
   
   The overall prefactors in \eqref{AGTblock} and \eqref{Zam} do not matter after taking four mass derivatives except for the $a$ dependent factors, which effectively replaces the $e^{-4\pi  \Im\tau_\text{UV}a^2}$ term in \eqref{MatrixModelSQCDSO8} by the $|16q|^{2a^2}$ factor in \eqref{MatrixModelSQCDSO82}. The other prefactors in \eqref{AGTblock} convert between the Nekrasov partition function of $SU(2)$ and $U(2)$, and only contribute to the quadratic in mass prefactor in \eqref{MatrixModelSQCDSO82} that vanishes after taking four derivatives.

\bibliographystyle{JHEP}
\bibliography{SQCD}

\end{document}